\pdfoutput=1
\documentclass[journal,transmag]{IEEEtran}
\makeatletter

\newcommand{\Rmnum}[1]{\expandafter\@slowromancap\romannumeral #1@}
\makeatother
\pdfoutput=1
\usepackage{cite}

\ifCLASSINFOpdf
   \usepackage[pdftex]{graphicx}
  \graphicspath{{./Figs/}{./jpg/}}
   \DeclareGraphicsExtensions{.pdf,.jpg,.png}
\else
   \usepackage[dvips]{graphicx}
   \graphicspath{{./Figs/}}
   \DeclareGraphicsExtensions{.eps}
\fi
\usepackage{amsmath}
\usepackage{bm}
\usepackage{amssymb}
\usepackage{gensymb}
\usepackage{flushend}
\usepackage{makecell}
\usepackage{siunitx}
\usepackage{fixltx2e}
\usepackage{xspace}
\usepackage{datetime}
\usepackage{doi}
\usepackage{algorithm,algorithmic}

\usepackage{multirow}
\usepackage{pbox}
\usepackage{array}

\ifCLASSOPTIONcompsoc
  \usepackage[caption=false,font=normalsize,labelfont=sf,textfont=sf]{subfig}
\else
 \usepackage[caption=false,font=footnotesize]{subfig}
\fi
\newcommand{\dif}{\mathop{}\!\mathrm{d}}

\begin{document}
	
\bstctlcite{IEEEexample:BSTcontrol}
%
\title{A Survey on STT-MRAM Testing:\linebreak Failure Mechanisms, Fault Models, and Tests}


\author{
	\IEEEauthorblockN{\makebox[\linewidth][c]}{Lizhou Wu\IEEEauthorrefmark{1}  {} {} {} Mottaqiallah Taouil\IEEEauthorrefmark{1} {}  {} {} Siddharth Rao\IEEEauthorrefmark{2} {} {}  {} Erik Jan Marinissen\IEEEauthorrefmark{2} {} {} {} Said Hamdioui\IEEEauthorrefmark{1}}
	\IEEEauthorblockA{ \IEEEauthorrefmark{1}Delft University of Technology {} {} {} {} {} {} {} {} {} {} {} {} {} {} {} {} {} {} {} {} {} {} {} {} {} {} {} {} {} {}  {} {} {} {} {} {} {} {} {} {} {} {} {} {} {} {} {} {} {} {} \IEEEauthorrefmark{2}IMEC {} {} {} {} {} {} {} {} {} {} {} {} {} {} {} {}  \\
		Mekelweg 4, 2628 CD Delft, The Netherlands  {}  {}  {} {}  {}  {} {}  {}  {}{}  {}  {} {}  {}  {} {} {} {} {} {} {} {}  Kapeldreef 75, B-3001 Leuven, Belgium {}  {}  {}
		\\\{Lizhou.Wu,  M.Taouil, S.Hamdioui\}@tudelft.nl  {}  {}  {} {} {} {} {} {} {} {} {} {} {}  {}  {}   \{Siddharth.Rao, Erik.Jan.Marinissen\}@imec.be {} {}  }
}

\markboth{A  Survey on STT-MRAM Testing, \today.}%
{Shell \MakeLowercase{\textit{et al.}}: Bare Demo of IEEEtran.cls for IEEE Transactions on Magnetics Journals}
%



\IEEEtitleabstractindextext{%
\begin{abstract}
 As one of the most promising emerging non-volatile memory (NVM) technologies, spin-transfer torque magnetic random access memory (STT-MRAM) has attracted significant research attention  due to several features such as   high density,  zero standby leakage, and nearly unlimited endurance. However, a high-quality test solution is required prior to the commercialization of STT-MRAM. In this paper, we present all STT-MRAM failure mechanisms: manufacturing defects, extreme process variations, magnetic coupling, STT-switching stochasticity, and thermal fluctuation. The resultant fault models including permanent faults and transient faults are  classified and discussed. Moreover, the limited test algorithms and design-for-testability (DfT) designs proposed in the literature are also covered. It is clear that test solutions for STT-MRAMs are far  from well established yet, especially when considering a defective part per billion (DPPB) level requirement.
 We present the main challenges on the STT-MRAM testing topic at three  levels: failure mechanisms, fault modeling, and test/DfT designs.
 

\end{abstract}

}

\maketitle

\IEEEdisplaynontitleabstractindextext

%
\IEEEpeerreviewmaketitle

\section{INTRODUCTION}
\label{sec:introduction}
Technology downscaling has driven a great success of the semiconductor industry in delivering faster, cheaper, and denser charge-based memories such as SRAM, DRAM, and Flash. However, as these existing memory technologies approach their scaling limits, they become increasingly power hungry and less reliable while the fabrication is more expensive due to the increased manufacturing complexity~\cite{Chen2017}.  
As alternative solutions, several promising non-volatile memory (NVM) technologies have emerged and attracted extensive R\&D attentions for various  levels in the memory hierarchy \cite{Chen2016,Yu2016,Xue2011}. Among them, spin-transfer torque magnetic random access memory (STT-MRAM) is considered as the leading candidate to replace SRAM for last-level caches (LLCs) in the short term and may serve as a universal memory technology in the long run \cite{Jog2012a}. The most attractive features of STT-MRAM, compared to SRAM and DRAM, are its non-volatility and nearly zero leakage power. The performance of STT-MRAM is customizable to the target application by having a trade-off between write latency and retention time \cite{Fong2016}.  Moreover, STT-MRAM offers an integration density ($6$$-$\SI{20}{F^2}) as high as DRAM ($6$$-$\SI{10}{F^2}) \cite{driskill2010latest},  essentially unlimited endurance ($>$$10^{15}$ cycles) \cite{Kan2016}, and   CMOS-compatibility.  Thanks to these attractive advantages, many companies worldwide have been heavily investing in the commercialization of STT-MRAMs.  For example, Everspin Technology announced the first STT-MRAM chip of 64Mb in 2012 \cite{slaughter2012high} and  the industry's  first 1Gb pMTJ-based STT-MRAM in 2016 \cite{EverSpin2016STT-MRAM}. Intel and Samsung  also demonstrated their embedded STT-MRAMs in 2018 \cite{Golonzka2018,SongYJ2018}. 

Despite the bright prospect of STT-MRAM technology,  an effective yet cost-efficient test solution is required before the mass production of  STT-MRAM chips. Manufacturing tests are responsible for weeding out all \textit{defective} chips prior to shipment to end customers. Therefore,  this is a very critical step in the entire VLSI design and manufacturing chain, since assembling a defective chip  onto a board or a  system causes  enormous cost and may even damage the manufacturer's reputation. 
STT-MRAM manufacturing process involves not only standard CMOS processing steps, but also the fabrication and integration of MTJ devices which are the data-storing elements.
The latter is subject to new manufacturing defects which have not been fully investigated so far.  Furthermore, due to the adoption of new materials and novel physical phenomena, new failure mechanisms such as magnetic coupling, STT switching stochasticity, and thermal fluctuation may cause transient faults, leading to yield loss \cite{Hamdioui2017}.  This shift in failure mechanisms may impact the fault modeling methodology and results. Note that accurate fault models which reflect the physical defects are crucial to develop a high defect coverage  test solution, e.g., \textit{defective part per billion} DPPB level.    Therefore, to develop a good-quality test, attention needs to be paid to the following three aspects: 1) understanding all failure mechanisms in STT-MRAMs so as to have accurate simulation models for them; 2) accurate fault analysis and  modeling; 3) test/design-for-testability (DfT) development to cover all faults.


This paper serves as a review of the-state-of-the-art on STT-MRAM testing. As this is still an emerging and ongoing research topic, we try to cover as much important related  work as possible in the literature. We organize and discuss the contents at three abstraction levels, which are failure mechanisms, faults models, and tests. First, we categorize all failure mechanisms for STT-MRAMs into five categories: 1) manufacturing defects; 2) extreme process variations; 3) magnetic coupling; 4) STT-switching stochasticity; 5) thermal fluctuation. Each of them will be introduced in details. Second, fault models due to those failure mechanisms are also classified and discussed in depth. The first two failure mechanisms result in permanent faults which are typically the targets of manufacturing tests. The other three, however, lead to transient faults which intermittently appear with certain occurrence rate in some specific cycles at run time. Thus, transient faults can be tolerated  by circuit-level techniques or error correction codes (ECCs), and they should be excluded somehow in any tests to avoid yield loss. Third, test algorithms and DfT designs proposed in the literature are also covered in this paper.  However, test solutions for  STT-MRAMs are still far away from established yet. To obtain a high-quality test, more research work needs to be done on this topic at failure-mechanism, fault-model, and test levels. 


The rest of this paper is organized as follows. Our discussion begins with an introduction to MRAM technologies with an emphasis on STT-MRAM in Section \ref{sec:mram_basics}. Section \ref{sec:defect_mechanisms} elaborates all failure mechanisms in STT-MRAM. Thereafter, we present fault models in Section \ref{sec:fault_models}.  Section \ref{sec:test_algorithms} and \ref{sec:dft}  discuss test algorithms and DfT designs, respectively.  The remaining challenges in STT-MRAM testing are presented in Section \ref{sec:discussion}. Finally, we conclude this paper  in Section \ref{sec:conclusion}.

\section{MRAM BASICS}
\label{sec:mram_basics}

\subsection{MTJ Fundamentals}
\label{subsec:MTJfundamentals}

\begin{figure}[!t]
	\centering
	\subfloat[Simplified MTJ structures with IMA (left) and PMA (right). ]{\includegraphics[width=0.3 \textwidth]{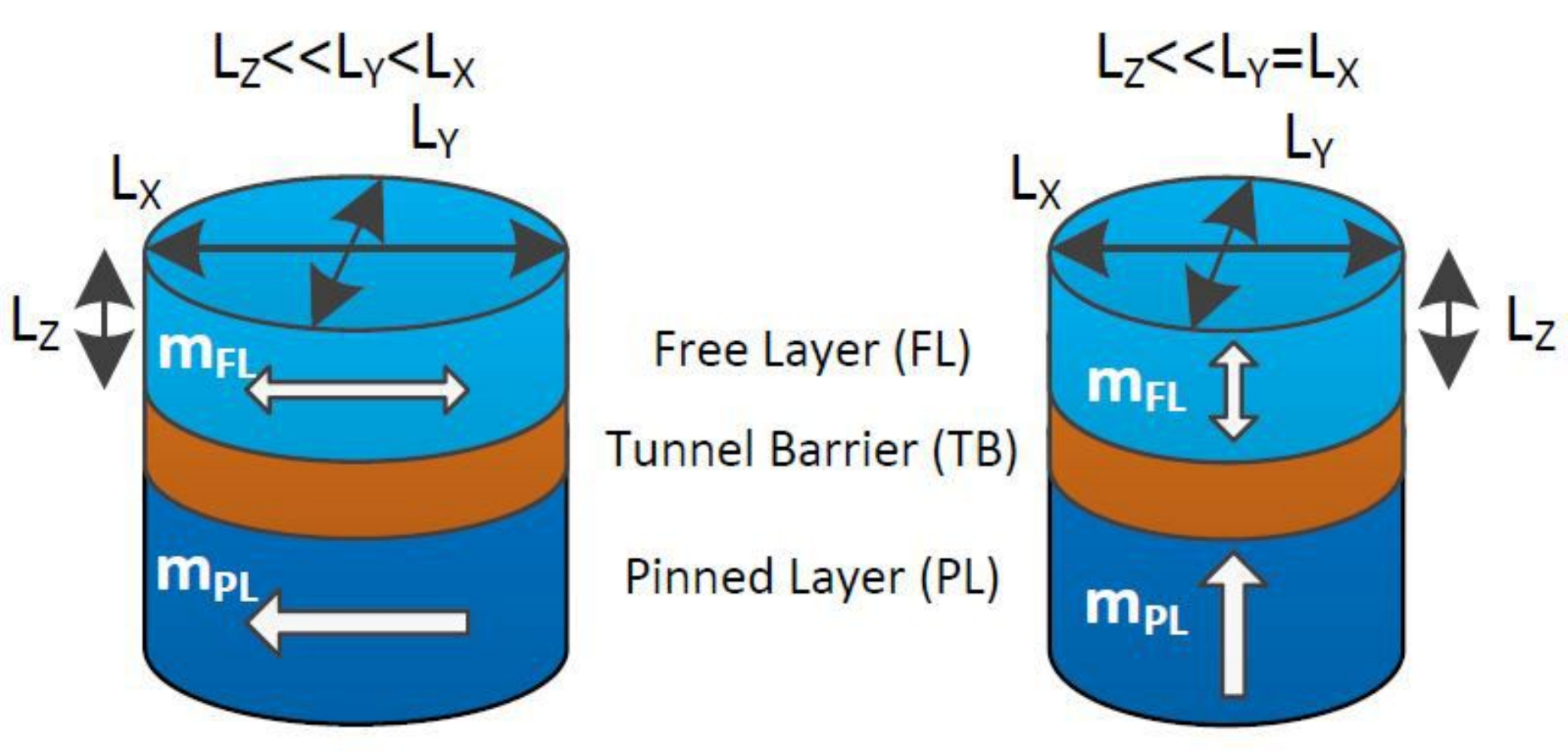}%
		\label{fig_MTJs}}
	\hfil
	\subfloat[Energy barrier ($E_\mathrm{B}$) between P and AP states.]{\includegraphics[width=0.4 \textwidth]{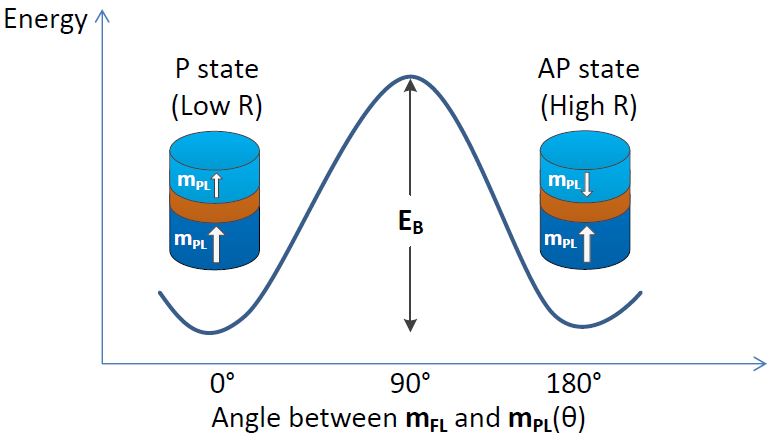}%
		\label{fig_EnergyBarrier}}
	\caption{Schematics of two basic MTJ structures and binary resistive states. }
	\label{fig_MTJ_basics}
\end{figure}

The \textit{magnetic tunnel junction} (MTJ) is the most basic building block for MRAMs; it essentially consists of two ferromagnetic layers sandwiching an extremely thin insulating spacer layer, as illustrated in Fig. \ref{fig_MTJ_basics}\subref{fig_MTJs}. The top ferromagnetic layer is called \textit{free layer} (FL) which stores the binary information. This layer is usually made of CoFeB material. The magnetization of the FL points along its intrinsic easy axis and may flip by applying a spin-polarized current through it. The MTJ device can be \textit{in-plane magnetic anisotropy} (IMA) \cite{huai2008spin,gajek2012spin,yuasa2004giant} if the easy axis points along x-axis (the left one in Fig. \ref{fig_MTJ_basics}\subref{fig_MTJs}), or \textit{perpendicular  magnetic anisotropy} (PMA) \cite{Ikeda2010,manchon2008analysis,nakayama2008spin,krounbi2015keynote} if along z-axis (the right one in Fig. \ref{fig_MTJ_basics}\subref{fig_MTJs}).
The bottom ferromagnetic layer, referred to as \textit{pinned layer} (PL) or \textit{reference layer}, is used to provide a stable reference direction to the magnetization in the FL. Although made of CoFeB as well, its anisotropy energy is large enough to avoid switching during operations.
The spacer layer in the middle is called \textit{tunnel barrier} (TB), which serves as an insulating non-magnetic spacer between the FL and PL. In case the TB layer is very thin (typically $\sim$\SI{1}{\nano\second}), quantum mechanical tunneling of electrons through the barrier makes the MTJ behave as a tunneling resistor, whose resistance depends exponentially on the barrier thickness. To evaluate the resistivity of MTJ devices, the \textit{resistance-area} ($\mathit{RA}$)  product is  commonly used in the MRAM community, as it is independent of the device size. Together, the above three layers form the fundamental structure of the MTJ device. 

The resistance of the MTJ device is low when the magnetization directions in FL and PL are parallel (P) and high when anti-parallel (AP), as shown in Fig. \ref{fig_MTJ_basics}\subref{fig_EnergyBarrier}. These two binary magnetic states enable the MTJ device to store one-bit data. 
The resistance difference between the P and AP states is attributed to the \textit{tunneling magneto-resistance} (TMR) effect \cite{Miyazaki2012,Yuasa2004, Khvalkovskiy2013}. The TMR effect can be simply interpreted by  the band model \cite{Khvalkovskiy2013} where  the good band matching in the P state leads to large tunneling conductance (i.e., low resistance), while the poor band matching in the AP state results in less electrons tunneling through the barrier (i.e., high resistance). To qualitatively evaluate the TMR effect, the $\mathit{TMR}$ ratio is widely adopted. It is defined by
\begin{equation}
\mathit{TMR}=\frac{R_{\mathrm{AP}}-R_\mathrm{P}}{R_\mathrm{P}} \times 100\% ,  \label{eq:TMR}
\end{equation}
where $R_{\mathrm{AP}}$ and $R_{\mathrm{P}}$ are the resistances in AP and P states, respectively. The higher the $\mathit{TMR}$, the easier it becomes for sense amplifiers to  distinguish the two magnetic states correctly (i.e., better readability). For commercially-feasible STT-MRAM products, a minimum $\mathit{TMR}$ ratio of $150\%$ is required \cite{Apalkov2016}. 

In order to switch between the AP and P states, a sufficient programming current is required to overcome the energy barrier ($E_\mathrm{B}$) between the two states, as shown in Fig. \ref{fig_MTJ_basics}\subref{fig_EnergyBarrier}.  $E_\mathrm{B}$ is given by \cite{Fong2016}:
\begin{equation}
E_\mathrm{B}=K_{\mathrm{u2}}V=\frac{\mu_0M_\mathrm{s}VH_\mathrm{k}}2,   \label{eq:E_B}
\end{equation}
where $K_{\mathrm{u2}}$ is the second-order uniaxial anisotropy constant, $V$ is the FL volume, $M_\mathrm{s}$ is the saturation magnetization, and $H_\mathrm{k}$  the magnetic anisotropy field. The energy barrier is key to the MTJ's \textit{thermal stability} ($\Delta$), which determines the retention time ($t_\mathrm{ret}$). The $\Delta$ and $t_\mathrm{ret}$ are expressed as \cite{Khvalkovskiy2013}:
\begin{equation}
\Delta= \frac {E_\mathrm{B}} {k_\mathrm{B}T},    \label{eq:Delta}
\end{equation}
\begin{equation}
t_\mathrm{ret}=t_0\exp(\Delta),       \label{eq:retention}
\end{equation}
where $k_B$ is  Boltzmann constant, and $T$ is the temperature. $t_0$ is the inverse attempt frequency ($\sim$1ns). The higher the $\Delta$, the more stable the magnetic state of the MTJ and thus the more energy required to program  it. For a nano-magnet  with $\Delta=40$, the retention time is around $7.4$ years \cite{strikos2013low}. Typically, $\Delta >80$ is needed to meet the industrial requirement, i.e., a retention time larger than 10 years.

There are fundamentally two kinds of MTJ devices that have been widely investigated. The first generation of MTJ devices is based on IMA, as illustrated with the left diagram in  Fig. \ref{fig_MTJ_basics}\subref{fig_MTJs}. Despite many early attempts of MRAM demonstration chips \cite{huai2004observation} or even commercial toggle-MRAM products by Freescale/Everspin in 2006 \cite{slaughter2005toggle}, IMA-MTJs have many shortcomings, including: 1) the elliptical cross-section with an aspect ratio of 2-3 makes them vulnerable to process variation, 2) decreased thermal stability with shrinking dimension, 3) further slash on the switching current is impractical at advanced technology nodes. As a result, the above limitations of IMA-MTJs have shifted research interest to  MTJ structures based on PMA. In PMA-MTJs, the uniaxial easy axis is perpendicular to the horizontal cross-section of MTJ (see the right side of Fig. \ref{fig_MTJ_basics}\subref{fig_MTJs} ). PMA-MTJs offer many benefits over IMA-MTJs, e.g., its circular structure ($L_Y=L_X$) makes it easier to scale to advanced nodes. Although there still exist many design challenges for PMA-MTJs, they are believed to be superior to the IMA-MTJs as building elements of STT-MRAMs at future technology nodes \cite{driskill2011latest}. Therefore, we limit our focus on STT-MRAMs based on PMA-MTJs in the remainder of this paper.

\subsection{ MRAM Classification}
\label{subsec:MRAMClassification}
In the past two decades, various MRAM technologies have been introduced. Despite their differences, a key distinction between them is the switching method of the MTJ magnetic state. Based on this, MRAM technologies can be classified into three categories: magnetic-field switching MRAM, STT switching MRAM, and novel-mechanism switching MRAM. Each of them is explained next.

\subsubsection{Magnetic-Field Switching MRAM}
The first MRAM generation is magnetic-field switching MRAM (MF-MRAM). The MF-MRAM uses an external magnetic field to switch the magnetic state \cite{slaughter2002fundamentals,Nahas2004}. Typically, a write operation to an MF-MRAM cell is implemented by injecting current to  metal lines above and below the addressed MTJ device. These current-carrying lines then generate magnetic fields to reverse the magnetization in FL. The biggest problem for MF-MRAM  is half-selection. It means that all non-targeted bit-cells along the two metal lines are exposed to the programming magnetic field (i.e., half-selected). This may cause inadvertent bit flips.

The half-selection problem can be mitigated by the Savtchenko switching technique \cite{savtchenko2003method}. This technique reduces the sensitivity of the half-selected cells by adopting: 1) synthetic antiferromagnet (SAF) structure for FL; 2) fixed sequence of write current pulse to toggle between AP and P states for the MTJ devices. MF-MRAM based on the Savtchenko switching technique is also called toggle MRAM \cite{engel20054}.

Despite several prototypes and  even small-scale commercial products \cite{slaughter2005toggle,EverSpin_Toggle_MRAM}, many limitations make mass production infeasible. For example, one of the challenges of MF-MRAMs is technology downscaling.  With technology shrinking, it is increasingly difficult to maintain a low write disturb rate and high thermal stability \cite{Apalkov2016}. Furthermore, the relatively high write current ($\sim$\SI{10}{\milli\ampere}) \cite {Cai2017} and  complicated cell geometry \cite{Apalkov2016} are also barriers to push MF-MRAMs to the market.

\subsubsection{STT Switching MRAM }

The spin-transfer-torque switching MRAM (STT-MRAM) \cite{berger1996emission,Slonczewski1996,huai2004observation,gajek2012spin} emerged as the second-generation of MRAM technology. It offers an alternative switching method that overcomes the scaling problems in MF-MRAMs. Due to the fact that STT-MRAMs leverage spin-polarized current to reverse the magnetization direction in FL, they have a much simpler memory cell geometry, eliminating the landing pads, word and bypass lines required in MF-MRAMs. This  makes it possible to scale the cell size of STT-MRAMs down to  4-6 \si{F^2}, compared to 20-30 \si{F^2} for MF-MRAMs. Therefore, research focus from both academia and industry has shifted from MF-MRAMs to STT-MRAMs for technology nodes of \SI{90}{\nano\meter} and below \cite{Apalkov2016}.

The STT switching process can be modeled by the Landau-Lifshitz-Gilbert (LLG) equation, which describes the magnetization dynamics of FL \cite{Slonczewski1996, Sun2000,aharoni2001micromagnetics}:

\begin{align}
  \frac{\dif{\bm m_{\mathrm{FL}}}}{\dif t} &=\bm\Gamma_{\mathrm{prec}}+\bm\Gamma_{\mathrm{damp}}+\bm\Gamma_{\mathrm{IP}}^{\mathrm{STT}}+\bm\Gamma_\mathrm{P}^{\mathrm{STT}}    \nonumber   \label{eq:LLGE}                   \\
  \bm\Gamma_{\mathrm{prec}}              &=-\bm\gamma\mu_0 \bm m \times \bm H   \nonumber                                \\
  \bm\Gamma_{\mathrm{damp}}              &=-\alpha\gamma \mu_0 \bm m \times (\bm m \times \bm H)         \\
  \bm\Gamma_{\mathrm{IP}}^{\mathrm{STT}}          &=\gamma\mu_0 \eta\frac \hbar2\frac Je\frac 1{M_\mathrm{s}t} \bm m_{\mathrm{FL}} \times (\bm m_{\mathrm{FL}} \times \bm m_{\mathrm{RL}})  \nonumber                                \\
  \bm\Gamma_{\mathrm{P}}^{\mathrm{STT}}           &=\gamma\mu_0 \eta\prime \frac \hbar2\frac Je\frac 1{M_\mathrm{s}t}   \bm m_{\mathrm{FL}} \times  \bm m_{\mathrm{RL}}.   \nonumber
\end{align}
In these equations, $\bm m_{\mathrm{FL}}= \frac{\bm M_{\mathrm{FL}}}{\bm M_\mathrm{s}}$ is the normalized  magnetization vector $\bm M_{\mathrm{FL}}$ in FL with respect to the saturation magnetization $\bm M_\mathrm{s}$, $\bm H$  the total effective magnetic field (including anisotropy and applied fields), $\alpha$ the Gilbert damping, $\gamma$  the gyromagnetic ratio, and $\mu_0$  the vacuum permeability, $J$ the current density, $t$ the thickness of FL, $e$ the electron charge, and $\hbar$ the Planck constant. The  term $\bm\Gamma_{\mathrm{prec}}$ in Equation (\ref{eq:LLGE}) describes the precessional motion of the magnetization around the effective field $\bm H$, while the  term $\bm\Gamma_{\mathrm{damp}}$ describes the gradual damping of the magnetization precessional motion toward the effective magnetic field $\bm H$. The term $\bm\Gamma_{\mathrm{IP}}^\mathrm{{STT}}$ and $\bm\Gamma_{\mathrm{P}}^{\mathrm{STT}}$ are the spin-polarized current induced torque acting on the $\bm m_\mathrm{{FL}}$, leading to a flip to the opposite direction.

By solving the LLG equation, the following expression for the critical switching current density ($J_{\mathrm{c0}}$) of the PMA-MTJs is derived \cite{Sun2000}:
\begin{equation}
  J_\mathrm{{c0}}^{\mathrm{PMA}}=\frac 1\eta\frac {2\alpha e}{\hbar}(M_\mathrm{s}t)(H_\mathrm{k}).      \label{eq:Jc0_PMA}
\end{equation}
This parameter is intrinsically determined by the MTJ design itself (e.g., materials, dimensions, and structure). This  makes it possible to compare the difficulty of switching the magnetic state (i.e., writability) among various MTJ designs. However, the actual switching process depends on both the amplitude and duration of the programming current in practical circuit designs \cite{koch2004time}. The higher the programing current density ($J_{\mathrm{c}}$) with respect to the $J_{\mathrm{c0}}$, the less the time required to complete a switching process. According to the duration of programming current pulse, two regimes are commonly observed to account for the switching process \cite{butler2012switching, bedau2010spin,myers2002thermally}. For very short pulses ($t_\mathrm{p}$$<$\SI{10}{\nano\second}), $J_c$  has to be much larger than the $J_{c0}$ to successfully switch the magnetization in FL. This is commonly referred to as the precessional regime, where the magnetization reversal is caused by STT effects. For longer pulses ($t_\mathrm{p}$$>$\SI{50}{\nano\second}), it is still possible that the magnetization in FL flips even if $J_{\mathrm{c}}<J_\mathrm{{c0}}$. This switching regime is called thermal activation regime, where the thermal fluctuation dominantly results in a magnetization flip. In practice, the programming current  lies in the precessional regime to reduce the write latency, while guaranteeing a determinative flip. Switching events in the thermal activation regime are one of the causes of undesired transient faults in STT-MRAMs.

Although STT-MRAM is considered as one of the most promising NVM technologies for both embedded and stand-alone applications in the future, it suffers from a few weaknesses. For example, the read disturb is non-negligible at advanced nodes, since writing and reading operations share the same path through the two-terminal MTJ device.  Another critical limitation is that the STT switching mechanism requires high currents ($>$\SI{1}{MA/cm^2}) with pulse widths in the \si{\nano\second} range \cite{VanBeek2017}. These programming currents may result in both hard and soft breakdown of the ultra-thin MgO barrier \cite{Khan2008,Yoshida2009, Schafers2009,Khan2015}.

\subsubsection{Novel-Mechanism Switching MRAM}
Serval new switching schemes  are under investigation as the third-generation MRAM, including spin hall effect (SHE), spin-orbit torque (SOT), voltage-control-magnetic anisotropy (VCMA), etc. The research motivation behind these novel  MRAM technologies is that the switching energy  can be further reduced, while boosting writing performance. The SHE effect utilizes spin current generated in the direction transverse to the charge current to switch the magnetization, which can be one order of magnitude less than the switching current in STT-MRAMs \cite {Kim2015a}. Since the effective spin injection efficiency can reach as high as 100\%, the SHE writing mechanism  has the potential to be faster ($>$1 GHz) and more energy efficient ($<$0.1 pJ/bit) \cite{KimYusung2013a}. Recently, the SOT switching mechanism has been introduced to MRAM. It offers an ultra-fast writing capacity with a higher reliability compared to the STT solution \cite{Kaushik2017,tsai2017spin}. Furthermore, The three-terminal setup of SOT-MTJ devices has separate read and write current paths, thereby solving the read disturb problem in STT-MRAM. The VCMA-MRAM is an another promising alternative to STT-MRAMs for low energy consumption applications \cite {Yoda2017}. It has the potential to simultaneously achieve ultrahigh storage density, ultralow energy consumption, and GHz high-speed operation at room temperature \cite{HuLi2012}. However, the uncontrollable random write-errors pose a severe reliability problem that needs to be overcome, since the VCMA only lowers the energy barrier between two magnetization polarizations.

In summary, three generations of MRAM technologies are under development based on the switching mechanism of MTJ magnetic state.  The first-generation MF-MRAM faces a number of challenges, especially the advanced scaling impeding its way to commercialization. The third-generation MRAM technologies are still in infancy stage and require further innovations in materials, reliability enhancement, manufacturing feasibility, etc. In contrast, STT-MRAM is currently the closest to wide deployment for a number of applications, such as consumer and industrial controllers, data centers, internet of things, and automotive \cite{MRAM-info}. We will limit our discussion to STT-MRAM in the remainder of this paper.

\begin{figure}[!t]
	\centering
	\includegraphics[width=0.5 \textwidth ]{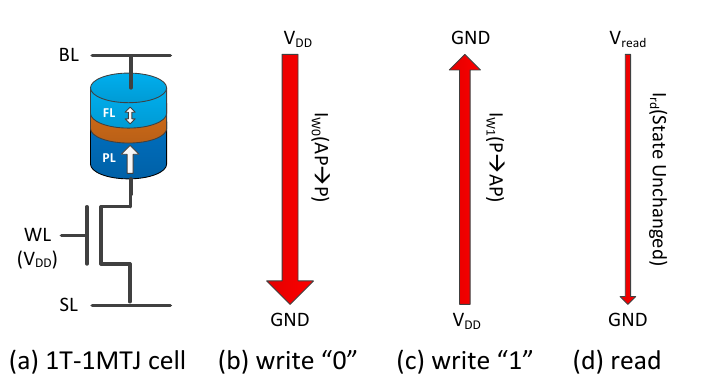}
	\caption{  Write and read operations on a 1T-1MTJ cell. }
	\label{fig_basicOps}
\end{figure}

\subsection{1T-1MTJ Bit-cell Design}
\label{subsec:bitcelldesign}
The 1T-1MTJ bit-cell design is the most widely-adopted cell design, comprising an MTJ device connected serially with an access transistor \cite{Lin2009,lee2010highly}, as shown in Fig. \ref{fig_basicOps}(a). The MTJ in this structure serves as a resistive storage element, while the access transistor, typically NMOS, is responsible for selective access. The NMOS gate is connected to a word line (WL), which determines whether a row is accessed or not.  The other two terminals are connected to bit line (BL) and source line (SL), respectively.  They control write and read operations on the internal MTJ device depending on the magnitude and polarization of voltage applied across them.

Fig. \ref{fig_basicOps}(b)-(d) show the three basic operations: write ``0'', write ``1'', and read. During a write ``0'' operation, WL and BL are pulled up to $V_{\mathrm{DD}}$ and  SL  is grounded, thus leading to a current ($I_{\mathrm{w0}}$) flowing from BL to SL. In contrast, a write ``1'' operation requires the opposite  current through the MTJ device with WL and SL at $V_{\mathrm{DD}}$, and BL grounded. In order to avoid write failures, write currents in both directions should be greater than the critical switching current $I_{\mathrm{c}}$. However, the current during a write ``1'' operation ($I_{\mathrm{w1}}$) is slightly smaller than during a  write ``0'' operation ($I_{\mathrm{w0}}$), due to the source degeneration of NMOS in write ``1'' operations \cite{Lee2012, Jones2012}. For read operations, a read voltage $V_{\mathrm{read}}$ is applied; it leads to a  read current ($I_{\mathrm{rd}}$) with the same direction as $I_{\mathrm{w0}}$ to sense the resistive state (AP/P) of MTJ.

To avoid an inadvertent state change during read operations, known as \textit{read disturb}, $I_{\mathrm{rd}}$ should be as small as possible; typically $I_{\mathrm{rd}}<0.5I_{\mathrm{c}}$ for MTJs with a thermal stability of $\Delta=60$ \cite{Zhao2011}. However, a too low $I_{\mathrm{rd}}$ may lead to \textit{incorrect read fault} \cite{Hamdioui2000}. In general, the current magnitude relations must satisfy: $I_{\mathrm{rd}}<I_{\mathrm{c}}<I_{\mathrm{w1}}<I_{\mathrm{w0}}$. This is indicated by  the widths of the red arrows  in Fig. \ref{fig_basicOps}. A read operation requires a sense amplifier to determine the resistive state. The sense amplifier may be implemented using a current sensing scheme, where the read-out value is determined by comparing the current of the accessed cell ($I_{\mathrm{cell}}=I_{\mathrm{rd}}$) with the current of a reference cell ($I_{\mathrm{ref}}$). The sensing result is logical ``0'' if $I_{\mathrm{cell}}<I_{\mathrm{ref}}$; otherwise, it outputs logical ``1''.

\pdfoutput=1

\begin{figure}[!t]
	\centering
	\subfloat[Bottom-up processing flow of STT-MRAM cells.]{\includegraphics[width=0.18 \textwidth]{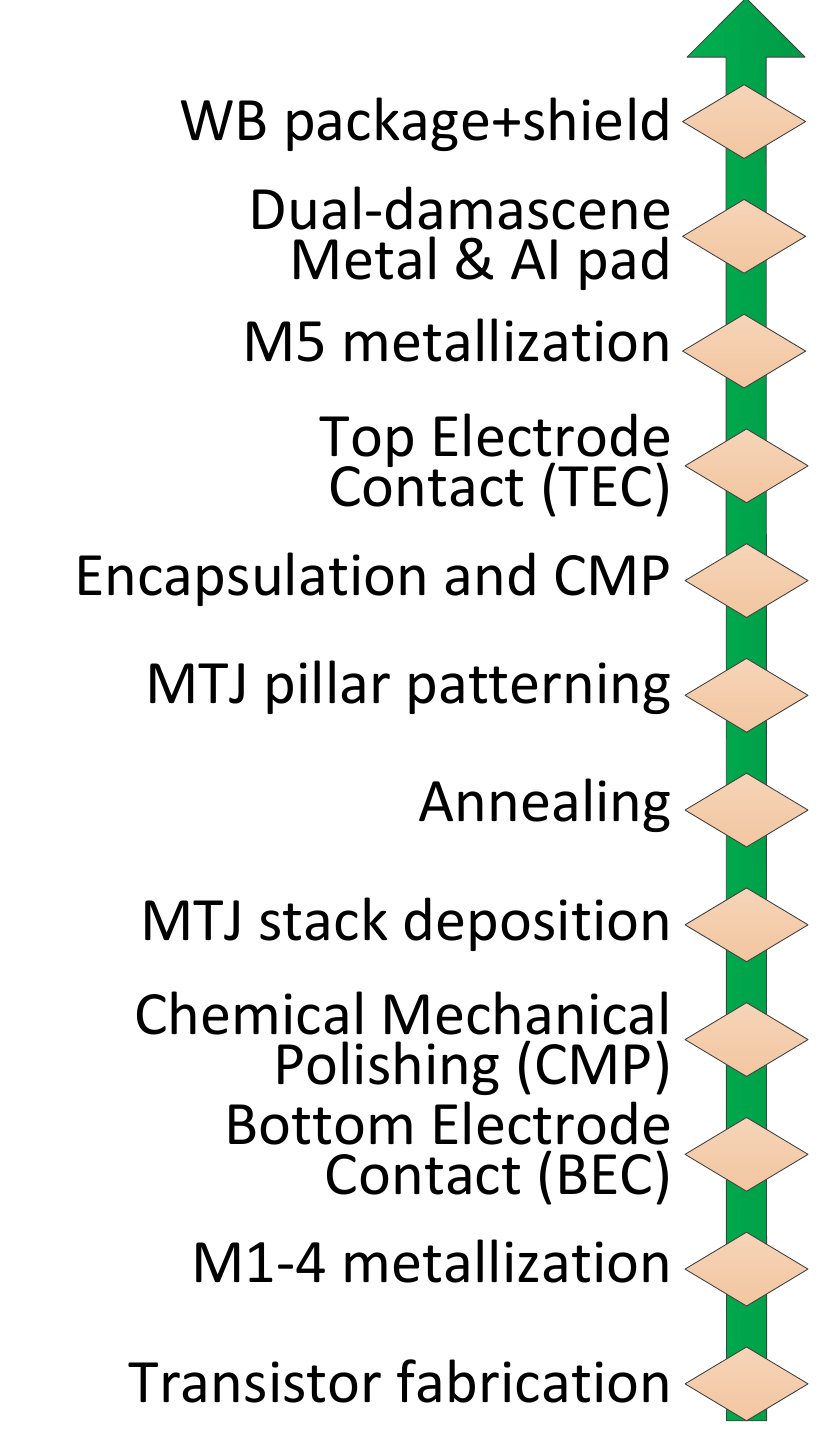}%
		\label{fig_MPFlow}}
	\hfil
	\subfloat[Vertical cross-section structure of  STT-MRAM cells from \cite{Song2016}.]{\includegraphics[width=0.27 \textwidth]{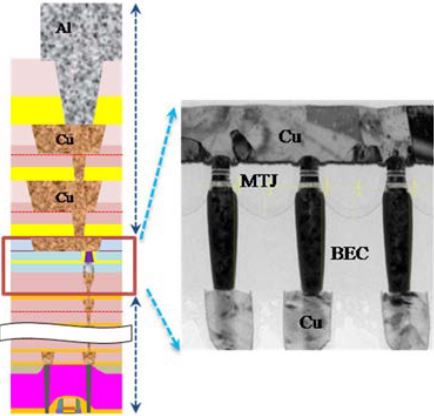}%
		\label{fig_1T1MTJcell}}
	
	\caption{General manufacturing process of STT-MRAM.}
	\label{fig_ManufacturingProcess}
\end{figure}

\section{FAILURE MECHANISMS}
\label{sec:defect_mechanisms}
Despite the small-scale manufacturability of STT-MRAM demonstrated by several semiconductor companies including Intel and Samsung,  STT-MRAM-specific failure mechanisms need to be fully studied and addressed before  mass production. First, the  fabrication of STT-MRAM chips requires a more sophisticated manufacturing process, including not only the mature CMOS fabrication steps, but also the MTJ fabrication and integration. The latter may also introduce new defects, which  unfortunately have not been fully investigated to date. Second, the  exploitation of magnetic materials and  novel physical phenomena makes STT-MRAMs suffer from some unique failure mechanisms including magnetic coupling, STT stochastic  switching, and thermal fluctuation. In this section, we  first discuss all potential manufacturing defects in STT-MRAMs. Thereafter, we  introduce the three STT-MRAM-specific failure mechanisms. 

\subsection {Manufacturing  Defects}
\label{subsec:manufacturingdefects}
 A defect is a physical imperfection in the processed wafer (i.e., an unintended difference from the intended design) \cite{Bushnell2000a}. To  guarantee  a  high-quality test solution as well as to improve the manufacturing process itself so as to improve yield,  understanding all potential defects  is of great importance.
The STT-MRAM manufacturing process mainly consists of the standard CMOS fabrication steps and  the integration of MTJ devices into  metal layers (e.g., between M4 and M5 layers \cite{Tillie2016,Shum2017}). Fig. \ref{fig_ManufacturingProcess}\subref{fig_MPFlow} shows the bottom-up processing flow and Fig. \ref{fig_ManufacturingProcess}\subref{fig_1T1MTJcell} the vertical cross-section structure of STT-MRAM cells.  Based on the manufacturing phase, STT-MRAM defects can be classified into front-end-of-line (FEOL)  and  back-end-of-line (BEOL) defects. As MTJs are integrated into metal layers during BEOL processing, BEOL defects can be further categorized into MTJ fabrication defects and metalization defects. All potential defects are listed in Table~\ref{table:PotentialDefects}. Next, we will examine  them in detail along with their corresponding processing steps, with a particular emphasis on those introduced during MTJ fabrication.

\begin{table}[!t]
	\caption{STT-MRAM defect classification.}
	\vspace*{-3mm}
	\scriptsize {
		\begin{tabular}{|l|l|l|}
			\hline\hline
			\multicolumn{1}{|c|}{FEOL}                   &                         \multicolumn{2}{c|}{ BEOL }                          \\ \hline
			\multicolumn{1}{|c|}{Transistor fabrication} & \multicolumn{1}{c|}{ MTJ fabrication} & \multicolumn{1}{c|}{ Metalization} \\ \hline\hline
			Material impurity                            & Pinholes in MgO barrier                & Open vias/contacts                  \\
			Crystal imperfection                         & Extreme thickness variation of TB      & Irregular shapes                    \\
			Pinholes in gate oxides                      & MgO/CoFeB interface roughness          & Big bubbles                         \\
			Shifting of dopants                          & Atom inter-diffusion                   & Small particles                     \\
			                                         & Redepositions on MTJ sidewalls         &                                 \\
			& Magnetic layer corrosion               &                                     \\
 \hline
		\end{tabular}
	}
	\label{table:PotentialDefects}
	\vspace*{-2mm}
\end{table}
\begin{figure}[!t]
	\centering
	\includegraphics[width=0.20 \textwidth ]{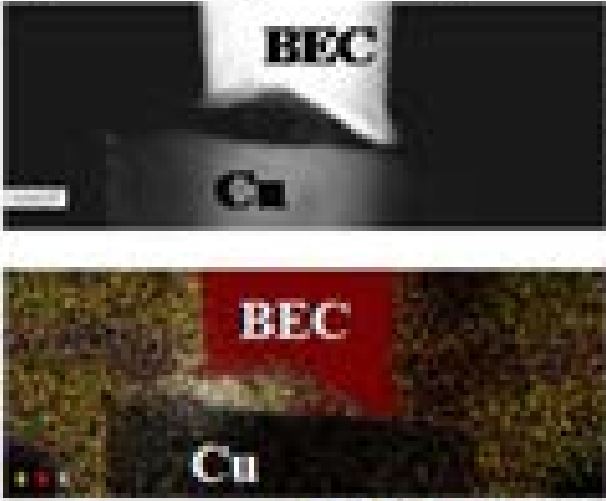}
	\caption{An open contact defect between the BEC and the underlying Cu layer. Reprinted from \cite{Song2016}.}
	\label{fig:resistive_open}
		\vspace{-10pt}
\end{figure}
\begin{figure*}[!t]
	\centering
	\subfloat[Schematic of a MTJ stack with a pinhole in the MgO tunnel barrier. ]{\includegraphics[width=0.5 \textwidth]{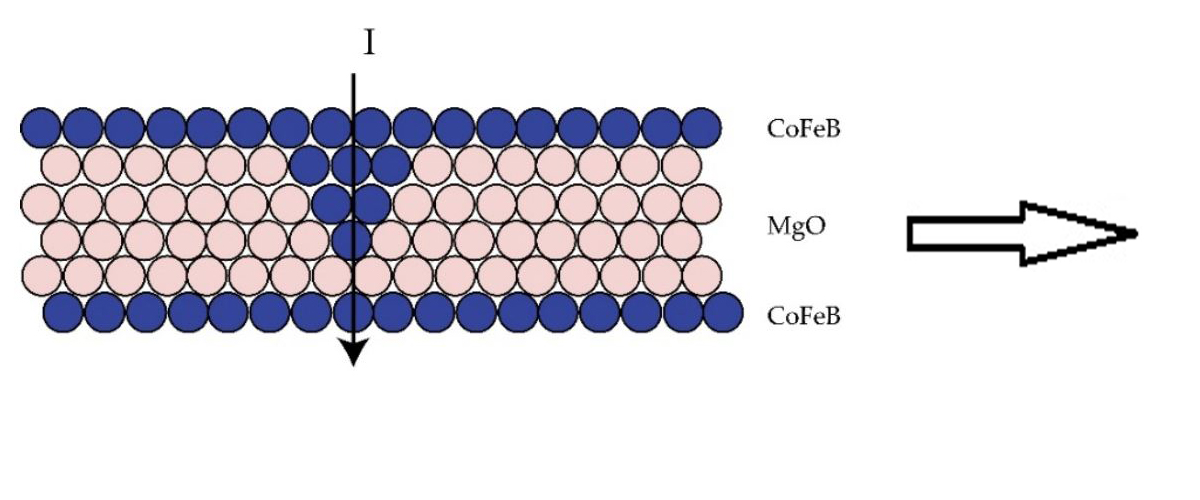}%
		\label{fig_PH}}
	\hfil
	\subfloat[Cross-section TEM  of a MTJ  with a pinhole defect.]{\includegraphics[width=0.35 \textwidth]{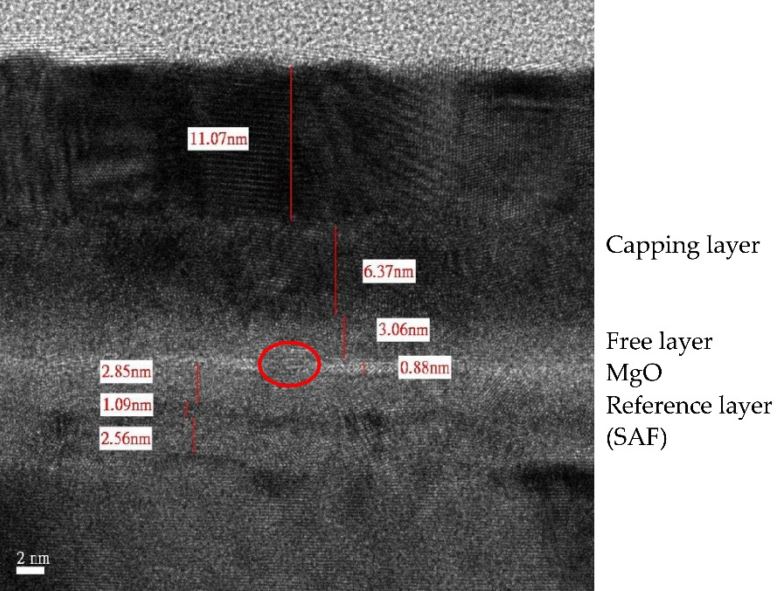}%
		\label{fig_MTJstack}}
	\caption{Pinhole defect in the tunnel barrier of  an MTJ device. Reprinted from \cite{Zhao2016}}
	\label{fig_pinhole}
\end{figure*}


\subsubsection{FEOL Defects}

The first step of the STT-MRAM manufacturing process is  the FEOL process where transistors are fabricated on the wafer. In this phase, typical defects may occur such as semiconductor impurities, crystal imperfections, pinholes in gate oxides, and shifting of dopants \cite{fedorenko2017ion,beenker1995defect}. These are the conventional  defects which have been sufficiently studied and are generally modeled by resistive opens, shorts and bridges  \cite{sachdev2007defect,li2005diagnosis,haron2011defect}.

\subsubsection{BEOL Defects}

After FEOL, M1-M4 metal layers  are stacked on top of the transistors followed by a bottom electrode contact (BEC), as illustrated in the zoomed-in part of Fig. \ref{fig_ManufacturingProcess}\subref{fig_1T1MTJcell}.
M1-M4 metalization does not differ from  traditional CMOS BEOL steps. The BEC step is used to connect bottom  Cu lines with MTJ stacks \cite{Kar2015,Song2016}. During this phase, typical interconnect defects may take place, such as open vias/contacts, irregular shapes, big bubbles, etc. \cite {sachdev2007defect}. Song et al. \cite{Song2016} provided a  transmission electron microscopy (TEM) image of an open contact, as shown in Fig.~\ref{fig:resistive_open}.  The open contact between the underlying Cu line and BEC is caused by polymer leftovers.

To obtain a  super-smooth interface between the BEC and the MTJ stack, a chemical mechanical polishing (CMP) step is required. The smoothness of the interface between layers is key to obtaining a good $\mathit{TMR}$ value. CMP processing minimizes the surface roughness with a root-mean-square (RMS)  average of 2\AA \cite{Tillie2016}. At this stage, both under-polishing and over-polishing of the surface can introduce defects.  Specifically, under-polishing  causes issues such as orange peel coupling or offset fields which affect the hysteresis curve, while over-polishing may result in dishing or  residual slurry particles that are left behind \cite{Vatajelu2017}.
\begin{figure}[!t]
	\centering
	\includegraphics[width=0.40 \textwidth ]{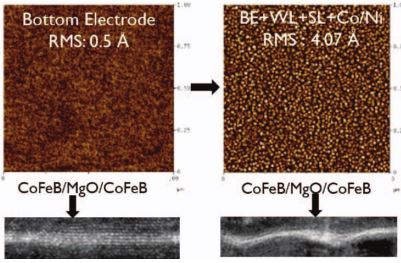}
	\caption{
	 The MgO/CoFeB interface is rougher for an advanced MTJ stack design (right) with an iSAF pinned layer than a simple MTJ stack design (left). Reprinted from \cite{Kar2015}.}
	\label{fig_Roughness}
\end{figure}

After the CMP step, the next critical step is to fabricate the MTJ stack. The latest published MTJ stack design consists of more than 10 films including a complicated inner synthetic anti-ferromagnetic (iSAF) pinned layer for performance reason \cite{ManuKomalan2017}.  However, the increasingly sophisticated MTJ design also makes it more vulnerable to manufacturing defects.  For example,  pinholes in the tunnel barrier (e.g., MgO) could be introduced in this phase. Fig. \ref{fig_pinhole}\subref{fig_PH} illustrates the concept of a pinhole defect, and Fig. \ref{fig_pinhole}\subref{fig_MTJstack} shows a vertical cross-section TEM image of a deposited MTJ stack with a pinhole in its \SI{0.88}{\nano\meter} tunnel barrier \cite{Zhao2016}. In this defective MTJ device, a pinhole forms in the tunnel barrier due to the rough deposition of MgO. As the CoFeB free layer is deposited on top of the tunnel barrier, the pinhole is filled with CoFeB material, as indicated by the red cycle in Fig. \ref{fig_pinhole}\subref{fig_MTJstack}.  Therefore, the pinhole filled with CoFeB material forms a defective high-conductance path across the two ferromagnetic layers. It severely degrades the resistance and TMR  value, and may even lead to breakdown due to the ohmic heating when an electric current passes through the barrier \cite{Oliver2004}. Furthermore, the MgO barrier thickness variation and interface roughness result in degradation of  resistance and TMR values as well. TEM images in \cite{Zhao2016} show  that the MgO barrier thickness varies from \SI{0.86}{\nano\meter} to \SI{1.07}{\nano\meter},  leading to a huge  difference in resistance.  Fig. \ref{fig_Roughness} shows with images of atomic force microscopy (top two)  and high-resolution transmission electron microscopy (bottom two)  that a complicated iSAF pinned layer design elevates interface roughness from \SI{0.5}{\AA} to \SI{4.07}{\AA}. The increased interface roughness leads to  significant TMR degradation \cite{Kar2015}.

\begin{figure}[!t]
	\centering
	\includegraphics[width=0.40 \textwidth ]{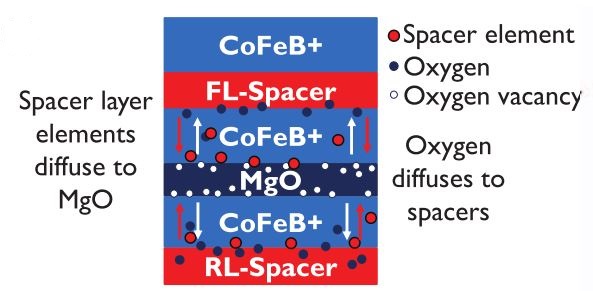}
	\caption{Schematic of atom inter-diffusion mechanism showing that oxygens diffuse out of the MgO barrier into  neighboring layers while spacer layer materials diffuse into the MgO layer. Reprinted from \cite{VanBeek2017}.}
	\label{fig_AtomInterdiffusion}
\end{figure}

\begin{figure*}[!t]
	\centering
	
	\subfloat[Schematic of ion beam etching with redeposition on the MTJ  sidewall. Reprinted from \cite{Zhao2016}. ]{\includegraphics[width=0.45 \textwidth]{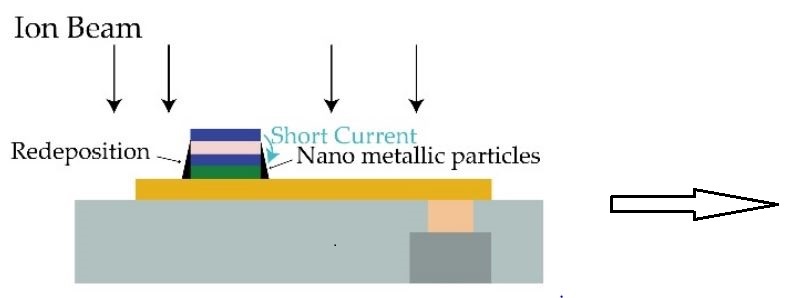}%
		\label{fig_Redeposition}}
	\hfil
	\subfloat[Vertical cross-section TEM images of an MTJ device with sidewall redeposition. Reprinted from  \cite{Sugiura2009}.]{\includegraphics[width=0.45 \textwidth]{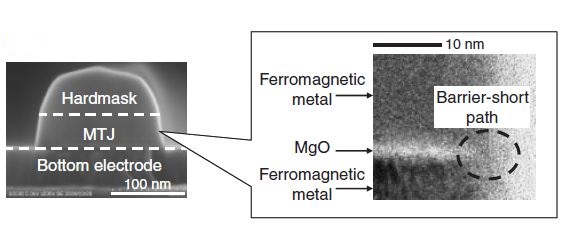}%
		\label{fig_MTJ_SEM1}}
	\caption{Magnetic material redeposition defect on the sidewall of MTJ devices.}
	\label{fig_MTJ_etching}
\end{figure*}

After the MTJ stack deposition, annealing is applied to obtain crystallization in MgO barrier as well as in CoFeB PL and FL layers \cite{meng2011annealing, maehara2011tunnel}. At this stage, the PMA originating from the MgO/CoFeB interface  and TMR value are strongly determined  by the  annealing conditions such as temperature, magnetic field and annealing time.  With appropriate annealing conditions, the PMA can be considerably enhanced, leading to higher thermal stability. Under-annealing can lead to lattice mismatch between the body-centered bubic (bcc)  CoFeB lattice and the fcc  MgO lattice, whereas over-annealing introduces atom inter-diffusion between layers. As illustrated in Fig. \ref{fig_AtomInterdiffusion}, oxygen atoms can diffuse out of MgO, leaving behind oxygen vacancies, thus severely degrading TMR value \cite{VanBeek2017}. Worse still, diffusion of Ta from the seed layer to MgO layer has been reported in several papers \cite{bhusan2012effect,Park2014a}, which scavenges O from MgO.

After MTJ multi-layer deposition, annealing and optical lithography processing, the next crucial step is to pattern individual MTJ nanopillars \cite{boullart2013stt}. Typically, Ion beam etching (IBE)  is widely used to pattern MTJ nanopillars \cite{Sugiura2009, Nagahara2006}. Fig. \ref{fig_MTJ_etching}\subref{fig_Redeposition} illustrates the etching process, where Ar ion beams are ionized and accelerated in a chamber and subsequently irradiate the wafer underneath, leading to selective etching of  the area where a hard mask does not cover. During the  MTJ etching process, it is extremely difficult to obtain desired MTJ nanopillars with steep sidewall edges, while avoiding sidewall redeposition and magnetic layer corrosion. The redeposition phenomenon on MTJ sidewall may significantly deteriorate the electrical property of the MTJ device, and even cause a barrier-short defect shown in Fig. \ref{fig_MTJ_etching}\subref{fig_MTJ_SEM1}. In order to mitigate the  redeposition effect, a side-etching step combined with the Halogen-based reactive ion etching (RIE) and inductively-coupled plasma (ICP) techniques \cite{nagahara2003magnetic, kim2012evolution, garay2015inductively} is needed by rotating and tilting the wafer. Nevertheless, other concerns arise. For instance, the shadowing effect (limited etching coverage at the lower corner of the MTJ profile due to insufficient spacing between MTJs) \cite{Zhao2016,Sugiura2009} limits a high-density array patterning, and magnetic layer corrosion degrades the reliability of MTJ devices due to the non-volatile chemicals attached to the CoFeB layers.

After MTJ etching processing,  encapsulation and CMP are required to separate individual MTJ pillars. In this step,  an oxygen showering post-treatment (OSP) can be applied to recover  patterning damage so as to improve the electric and magnetic properties of MTJ devices \cite{Jeong2015a}.   The oxygen showering process selectively oxidizes the perimeter (damaged by  previous ion beam etching) of the MTJ pillar with non-reactive oxygen ions.  However, over-oxidization too much into the MTJ device also causes degradation in key device parameters such as TMR. Thus, the OSP condition needs to be carefully tuned to maximize the damage suppression while protecting the inner undamaged parts. 

Next, MTJ pillars are connected to top electrode contact (TEC), followed by M5 metallization. The rest of manufacturing process are same as the BEOL steps of CMOS technology. Typical defects such open contact/vias, small particles etc. can occur in this phase as well. It is worth-noting that a package-level magnetic shield can be added to enhance the stand-by magnetic immunity of STT-MRAMs, as proposed in \cite{Lee2018}. The magnetic shield was reported to be effective in protecting STT-MRAMs against  external magnetic fields.

\subsection {Extreme Process Variations}
\label{subsec:extremePV}
Apart from manufacturing defects and magnetic coupling induced defects that we discussed in the previous section, extreme process variations are another probable cause of permanent faults in STT-MRAMs \cite{Kang2015a, Chintaluri2016a}.
Process variations  can be introduced at  each step of the STT-MRAM manufacturing process, leading to parasitic variations in both MTJs and transistors.   Some key MTJ parameters affected significantly by process variations are listed as follows \cite{Chintaluri2016a, Kang2015}:

\begin{itemize}
	\item  Magnetic anisotropy ($H_\mathrm{k}$)
	
	\item  Saturation magnetization ($M_\mathrm{s}$)
	
	\item  Tunnel magnetoresistance ratio ($\mathit{TMR}$)
	
	\item  Tunnel barrier thickness ($t_{\mathrm{ox}}$)
	
	\item  Cross-sectional area  ($A$)
\end{itemize}
Similarly, the fabrication of transistors also introduces cell-to-cell variations in some key parameters, such as the threshold voltage ($V_{\mathrm{th}}$) and the transistor size ($L\times W$), caused by random dopant fluctuations and line-edge roughness, etc. \cite{ye2011statistical, zhao2007predictive}. As CMOS transistors not only reside in memory cells (serving as access controllers)  but also in peripheral circuits,  process variations of transistors also have a detrimental impact on  STT-MRAMs. Process  variations in the access NMOSs of bit-cells mainly affect their current driving capabilities during write/read operations, whereas process variations of transistors in peripheral circuits result in reliability degradation of write and read operations (e.g., sensing margin decrease).

All these parametric variations in both MTJs and transistors pose a huge threat to STT-MRAM designs. In practice, the worst case has to be considered and enough guard bands must be provided in designs for a target failure probability, typically $\sim$$10^{-9}$ for a $6\sigma$ corner  \cite{Chintaluri2016a}. However, the band-guard scheme also results in serious performance sacrifice. The authors in \cite{Chintaluri2016a} claimed that the $6\sigma$-corner values of write latency are 3.5x larger than the mean value for a given thermal stability  $\Delta=60$. Furthermore, this guard-band scheme has been afflicted by another notorious drawback, which is  energy waste for the majority of cells in an STT-MRAM array. Under the circumstance of mass production, those cells with around or over $6\sigma$  parameter deviation are inevitable to suffer from permanent faults as follows.

\subsection {Magnetic Coupling}
\label{subsec:magneticcoupling}
As STT-MRAMs store data as  relative orientation of magnetizations in the FL and PL of MTJ devices, the stability of magnetic states in these two ferromagnetic layers is vulnerable to any extra unintended magnetic fields  externally or internally. As introduced in Section \ref{subsec:MTJfundamentals}, an MTJ device is composed of multiple ferromagnetic layers, which  all inevitably generate \textit{stray fields} ($H_{\mathrm{stray}}$) in the space \cite{Wang2012}. These stray fields in turn have an impact on the stability of magnetization in the FL \cite{Augustine2010, Bandiera2010,Xue2014a,Insik_ITC_2016}.    $H_{\mathrm{stray}}$ varies with the MTJ ferromagnetic materials, stack design, dimensions, spacing, and process variation from device to device.  Apart from the intra-cell stray fields, the ferromagnetic layers from neighboring cells also generate stray fields \cite{Yoon2018}. Therefore, all these intra-cell and inter-cell stray fields together can form a net offset field ($H_{\mathrm{offset}}$)   on the FL of a specific MTJ within an STT-MRAM array.  The effect of $H_{\mathrm{offset}}$ on a victim cell is usually referred to as \textit{magnetic coupling}.  As technology scales down, STT-MRM cells  move even closer to each other; this further aggravates the magnetic coupling problem.  Thus, it is of great importance to investigate the effect of magnetic coupling on the STT-MRAM performance and reliability.

\begin{figure}[!t]
	\centering
	\includegraphics[width=0.4 \textwidth ]{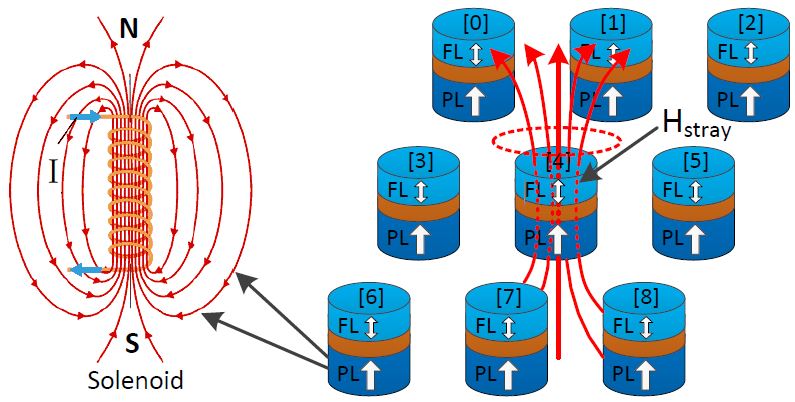}
	\caption{Schematic of magnetic coupling. Neighboring cells 0-3 and 5-8 (aggressors) together generate an unintended  stray field $H_{\mathrm{stray}}$ at the central cell 4 (victim). $H_{\mathrm{stray}}$ always exists during the lifetime of STT-MRAMs with its magnitude varying with the data pattern in the neighborhood.}
	\label{fig_MagneticCoupling}
\end{figure}
\begin{figure}[!t]
	\centering
	\includegraphics[width=0.43 \textwidth ]{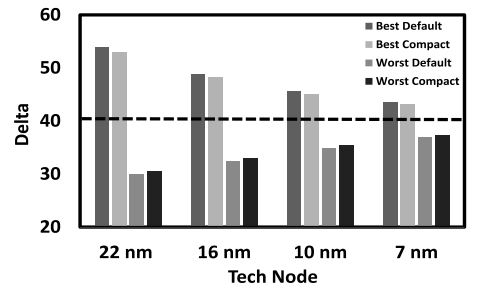}
	\caption{$\Delta$ variation (nominal $\Delta=40$) due to magnetic coupling with respect to various technology nodes for PMA-MTJs. The spacing  in  the default cell array is 5F, whereas it is 3F in the compact array. Reprinted from \cite{Yoon2018}.  }
	\label{fig_Delta_MC}
\end{figure}

Fig. \ref{fig_MagneticCoupling} shows a 3$\times$3 PMA-MTJ  array. All ferromagnetic layers (i.e., FLs and PLs) of MTJs in the neighborhood of MTJ 4 together generate a net offset field $H_{\mathrm{offset}}$ acting on the victim MTJ 4 in the center. The stray field of a single ferromagnetic layer can be modeled as  the magnetic field of a solenoid, shown in Fig. \ref{fig_MagneticCoupling}. The amount of current for the solenoid  to generate the same amount of magnetization in any ferromagnetic layer can be calculated by $\frac {M_{\mathrm s}t}{N}$ \cite{Yoon2018}, where $M_{\mathrm{s}}$ is the saturation magnetization, $t$ is the thickness of the ferromagnetic layer, and $N$ is the number of coils. By means of this solenoid model, we can approximately calculate $H_{\mathrm{stray}}$  of each ferromagnetic layer at any spot in space by the Biot-Savart law. Therefore,  $H_{\mathrm{offset}}$ at the  FL point in the  victim MTJ 4 can be derived for a given data pattern in the 3$\times$3 MTJ array.

The magnetic coupling effect has an impact on various MTJ parameters. Firstly, the thermal stability $\Delta$ of the MTJ  may deviate from its nominal value either positively or negatively depending on the direction of $H_{\mathrm{offset}}$ with respect to the anisotropy field $H_{\mathrm{k}}$ of the MTJ's FL. The effect of $H_{\mathrm{offset}}$ on  $\Delta$ can be described by \cite{Khvalkovskiy2013}
\begin{equation}
\Delta(H_{\mathrm{offset}})=\Delta(H_{\mathrm{offset}}=0) (1\pm\frac {H_{\mathrm{offset}}}{H_{\mathrm{k}}})^2 \label{eq:Delta_offsetfield}.
\end{equation}
For example,  $H_{\mathrm{offset}}$ reaches its peak when the data pattern is [111$x$,1111], indicating that all the neighboring MTJs are in AP state \cite{Yoon2018}. In this case, the thermal stability $\Delta$ of the central victim MTJ 4 is enhanced to the best extent if it is in P state (i.e., $x=0$). Conversely, the worst data pattern [1111,1111] weakens the $\Delta$ value  to its rock bottom.
Fig. \ref{fig_Delta_MC} shows the $\Delta$ values under the best data pattern and the  worst data pattern with regard to various technology nodes. The gap between the best and worst cases is slightly over $50\%$ of the nominal  $\Delta=40$ at \SI{22}{\nano\meter} node, and it shrinks with technology scaling down. This is because $H_{\mathrm{k}}$ increases faster than $H_{\mathrm{offset}}$ in order to maintain the  nominal  $\Delta$ at 40 for various technology nodes \cite{Yoon2018}.

Secondly, the magnetic coupling effect also influences the critical switching current density $J_{\mathrm{c0}}$. The effect of magnetic coupling on $J_{\mathrm{c0}}$  can be characterized by adding a $H_{\mathrm{offset}}$ term to Equation \eqref{eq:Jc0_PMA} \cite{Khvalkovskiy2013}:
\begin{equation}
  J_{\mathrm{c0}}^{\mathrm{PMA}}=\frac 1\eta\frac {2\alpha e \mu_0}{\hbar}(tM_{\mathrm{s}}H_{\mathrm{k}})(1\pm\frac{H_{\mathrm{offset}}}{H_{\mathrm{k}}})     \label{eq:Jc0_mc}.
\end{equation}
As an MTJ device with a higher $\Delta$ requires a larger write current to switch the magnetization in the  FL, the best data pattern for $\Delta$ now becomes the worst case for $J_{\mathrm{c0}}$. Similarly, the worst data pattern for $\Delta$ becomes the best case for $J_{\mathrm{c0}}$, meaning that a smaller write current is required to reverse the  magnetization in the FL. Consequently,  $J_{\mathrm{c0}}$ variation due to magnetic coupling amplifies the stochasticity of the STT switching behavior, thus necessitating a large write margin to maintain an acceptable write error rate for all memory cells.
\begin{figure}[!t]
	\centering
	\includegraphics[width=0.4 \textwidth ]{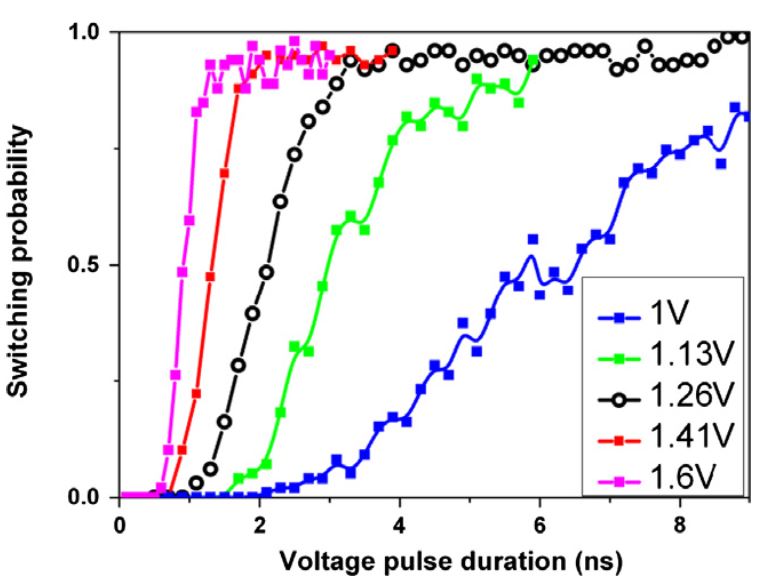}
	\caption{Experimental measurements of STT-induced switching probability vs.  duration and  amplitude of the applied voltage pulse.  Reprinted from \cite{Zhao2012a}. }
	\label{fig_switchingP}
\end{figure}

\subsection {STT-switching Stochasticity}
\label{subsec:STTStochasticswitching}
In Section \ref{subsec:MRAMClassification}, we have briefly introduced the STT-induced switching mechanism. For a write current pulse  shorter than \SI{10}{\nano\second}, the STT-effect dominantly results in a magnetization switching action if the density of the spin-polarized current  flowing through the FL of MTJ device is larger than the critical switching current density ($J_c$$>$$J_{c0}$).  The LLG equation (\ref{eq:LLGE}) models the magnetization dynamics of the precessional switching process. However, the actual switching time varies from one event to the next, due to the fact that the STT-switching is intrinsically stochastic \cite{Zhao2012a}.   According to the experiments and theoretical analysis in \cite{Devolder2008}, after the write pulse onset there exists a ns-scale incubation delay time varying significantly for individual events. Then, it is followed by an  abrupt precessional switching of the magnetization for around \SI{400}{\pico\second}. Therefore, a switching failure occurs when the incubation time gets longer than the fixed pulse width of  write operations.

Fig. \ref{fig_switchingP} shows the switching probability of the magnetization in the FL versus the amplitude and duration of the applied voltage pulse. It can be seen that the switching probability increases remarkably with the pulse duration and amplitude before reaching an saturation level close to 100\%.  Thus, increasing the write current amplitude or duration is an effective method to avoid write failure \cite{Zhao2012a, MarinsDeCastro2012}.  However, this approach also leads to significant power and speed overhead, which has posed a major obstacle to the potential application of STT-MRAMs as last level caches (LLCs).

\begin{figure}[!t]
	\centering
	\includegraphics[width=0.43 \textwidth ]{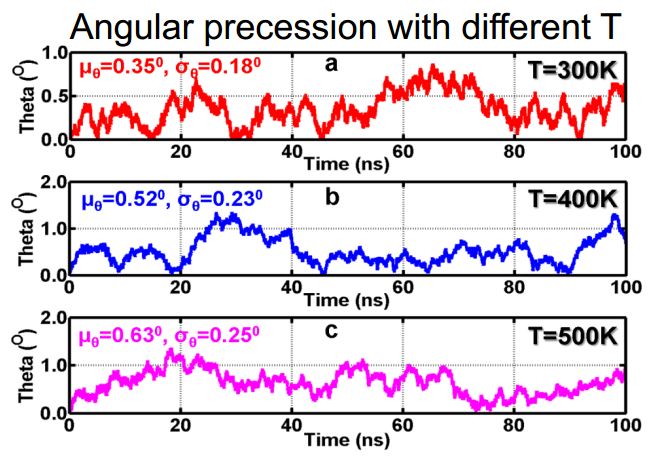}
	\caption{Thermally-induced initial angle oscillation over time under the ambient temperature of 300K, 400K, and 500K. Reprinted from \cite{Roy2014s}. }
	\label{fig_thermalFluctuation}
\end{figure}
\begin{figure*}[!t]
	\centering
	\subfloat[The temperature dependence of  data lifetime  in MTJ devices. Reprinted from \cite{Roy2014s}.]{\includegraphics[width=0.4 \textwidth]{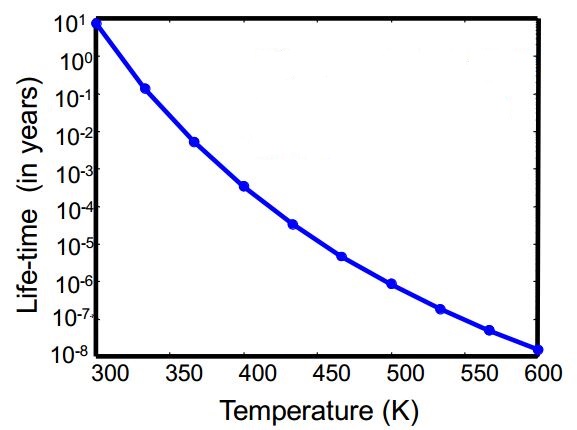}%
		\label{fig_retentionT}}
	\hfil
	\subfloat[The temperature dependence of TMR value and switching current. Reprinted from \cite{BiWu2016}.]{\includegraphics[width=0.5 \textwidth]{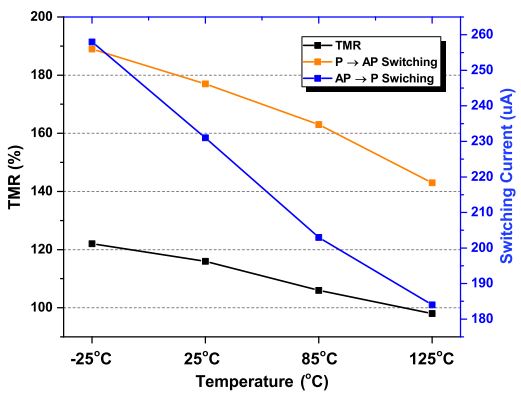}%
		\label{fig_TMR_Temperature}}
	\caption{The ambient temperature has a  evident impact on MTJ parameters: (a) thermal stability (manifested as the lifetime of the stored data in MTJ devices), (b) TMR value and switching current. }
	\label{fig_TMR_Reten_T}
\end{figure*}

\subsection {Thermal Fluctuation}
\label{subsec:thermal_fluctuation}
Thermal fluctuation has a great impact on the STT-switching behavior; it increases the cycle-to-cycle magnetization switching variation \cite{Wang2008}.  The effect of thermal fluctuation on the STT-switching behavior can be characterized by modifying the LLG Equation (\ref{eq:LLGE}), taking into consideration the thermally-induced random field ($\bm H_{\mathrm{fluc}}$) and the initial angle ($\theta$) between the magnetizations in the FL and the PL \cite{Kang2015,sankey2008measurement,Kang2016a}:
\begin{align}
\frac{\dif{\bm m_{\mathrm{FL}}}}{\dif t} &=\bm\Gamma_{\mathrm{prec}}+\bm\Gamma_{\mathrm{damp}}+\bm J(\theta)(\bm m \times \bm m \times \bm m_{\mathrm{RL}})      \label{eq:LLG_ThermalFluc}                   \\
\bm\Gamma_{\mathrm{prec}}              &=-\bm\gamma\mu_0 \bm m \times (\bm H + \bm H_{\mathrm{fluc}})  \nonumber                                \\
\bm\Gamma_{\mathrm{damp}}              &=-\alpha\gamma \mu_0 \bm m \times (\bm m \times (\bm H + \bm H_{\mathrm{fluc}}), \nonumber
\end{align}
where $\bm J(\theta)$ is the coefficient of the STT term, depending on the initial angle ($\theta$) between the magnetizations in the FL and the PL. Based on the above equation, Wang et al. \cite{Wang2008} claimed that thermal fluctuation not only influences the STT-switching time distribution, but also plays a major role in determining the magnetization reversal for a long pulse width (typically $>$\SI{100}{\nano\second}) in the thermal activation regime. In other words, a small current even lower than the critical switching current density ($J_c$$<$$J_{c0}$) is still possible to reverse the magnetization in the FL.  This situation may occur in read operations, thus causing a read disturb fault.  Worse still, an unexpected magnetization flip may happen even if no current flows through the MTJ device, leading to a retention fault. For the  thermally-induced switching under a long pulse, the switching probability can be estimated by the Neel-Brown model \cite{Louis2009}
\begin{align}
Pr(t)=1-\exp(-t/\tau_1)                 \label{eq:readdistrub}\\
\tau_1=\tau_0 \exp(\Delta(1-I/I_{\mathrm{c0}})),    \nonumber
\end{align}
where $\tau_0$ is the attempt time ($\sim$\SI{1}{\nano\second}) characterizing the timescale under which the magnetization can be considered practically at rest, and $\tau_1$ represents the averaged switching time. $I$ and $t$ are the amplitude and duration of the applied current, respectively. $I_{c0}$ is the critical switching current.

Furthermore, the  thermal fluctuation magnitude significantly depends on the ambient temperature. The effect of temperature is twofold. First, thermal fluctuation agitates a random initial angle ($\theta$) between the magnetizations in the FL and the PL, and both the mean value and the standard deviation  $\theta$ increase with temperature \cite{Augustine2011,Khvalkovskiy2013}.  Fig. \ref{fig_thermalFluctuation} shows that the oscillation intensity of $\theta$ is nearly doubled as the temperature goes from $300K$ to $400K$. This makes STT-MRAMs subject to severe reliability problems at elevated temperature. Second, temperature has a significant influence on some MTJ parameters, such as $\Delta$, TMR, and switching current.   According to equation (2-4), since $\Delta$ is inversely proportional to temperature, the lifetime of  data stored in STT-MRAM cells decreases exponentially with temperature, as shown in Fig. \ref{fig_TMR_Reten_T}\subref{fig_retentionT}.  Zhao et al. found that $R_{\mathrm{AP}}$ decreases while $ R_{\mathrm{P}}$ barely changes  when the temperature rises \cite{Kou2006,zhao2012spin,Dqj2016}. As a result, TMR value goes down with temperature, as shown in Fig. \ref{fig_TMR_Reten_T}\subref{fig_TMR_Temperature}. This makes read margin shrink at elevated temperature, which may  lead to an  incorrect read fault. Apart from $\Delta$ and TMR, the switching current also decreases with temperature \cite{lee2010design}. However, this is actually a benefit for write operations, meaning that a shorter or smaller write pulse is required at higher temperature. By utilizing this phenomenon, thermally-assisted MRAMs have been proposed to briefly heat up the MTJ device in write operations to facilitate the magnetization reversal \cite{prejbeanu2007thermally,chaudhuri2010design}.

\pdfoutput=1
\section{FAULT MODELS}
\label{sec:fault_models}
In the last section, we have introduced  all potential defects that may take place in STT-MRAMs. The occurrence  rate of these defects  mainly depends on the manufacturing technology and process. To detect those defective STT-MRAM chips,   it is  prohibitively expensive to  physically and manually examine  all manufactured chips. As an alternative, it is usually more efficient and cheaper to characterize and detect faulty behaviors  from a functional perspective. This calls for accurate STT-MRAM fault models which are the representation of certain physical defects at the abstracted function level \cite{Bushnell2000a}. Despite the well-established fault models for traditional memory technologies, STT-MRAM, as an emerging NVM technology, may need unique fault models to cover all potential defects. In the section, we will present all proposed fault models in the literature. 

\begin{figure}[!t]
	\centering
	\includegraphics[width=0.45 \textwidth ]{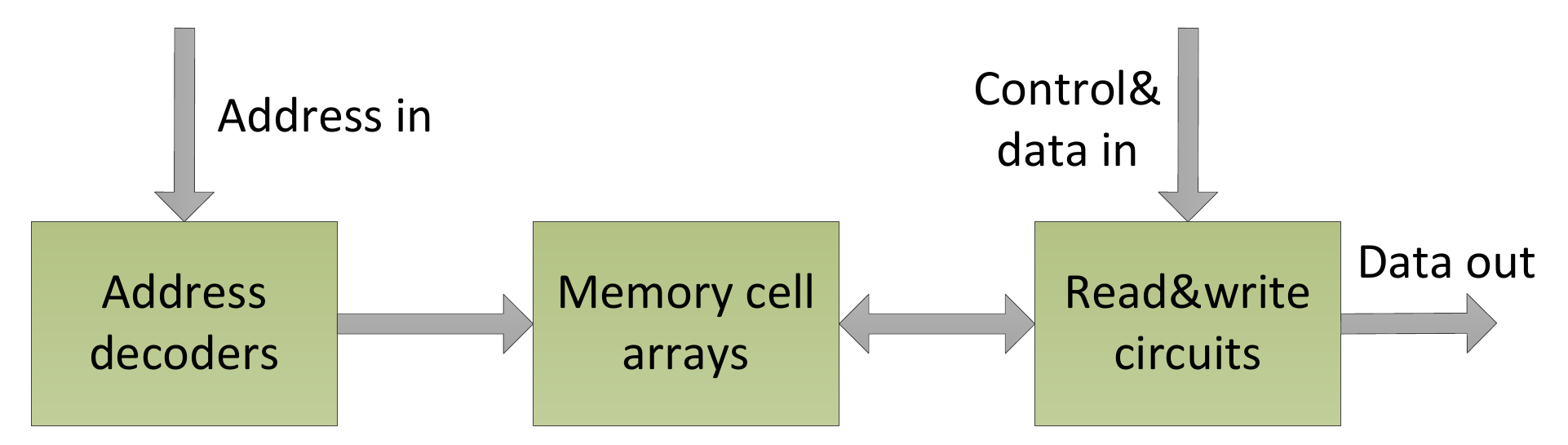}
	\caption{Reduced functional memory model.}
	\label{fig_functionalblock}
\end{figure}

\begin{figure}[!t]
	\centering
	\includegraphics[width=0.5 \textwidth ]{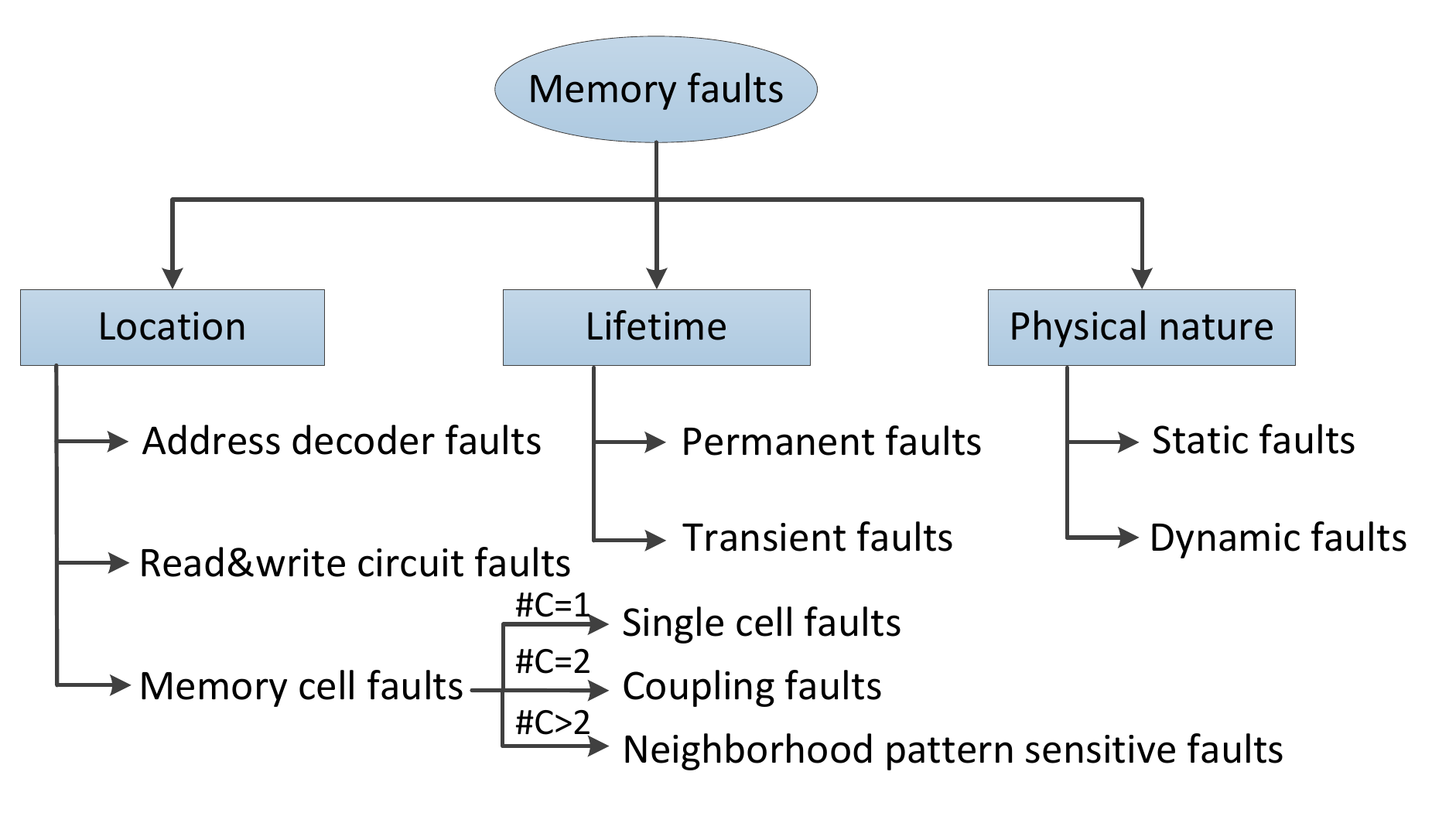}
	\caption{Classification of memory faults.}
	\label{fig_faultclassification}
\end{figure}

\subsection{Classifications}
\label{subsec:fault_classification}
A memory chip can be functionally reduced to three blocks: address decoders, memory cell arrays, and read\&write circuits, as shown in Fig. \ref{fig_functionalblock}. Address decoders are concerned with addressing the right cell or word in a memory cell array.  Memory cell arrays are composed of identical memory cells in a matrix form. Read\&write circuits usually consist of sense amplifiers, pre-charge circuits, and write drivers; they are responsible for data transport from and to the addressed memory cell, respectively.

Fig. \ref{fig_faultclassification} illustrates the classification of memory faults based on three criteria:  1) location,  2) lifetime, and 3) physical nature. We will explain each of them as follows.
\subsubsection{Location}
According to the fault location, faults can be grouped into three categories: address decoder faults, memory cell array faults, and read\&write circuit faults. Address decoder faults are concerned with those faults in the address decoders, such as \textit{mismatch faults} between addresses and cells, \textit{delay faults} due to the increased capacitance in word lines or bit lines, etc. Memory cell array faults refer to those faults taking place in memory cell arrays, such as \textit{stuck-at-fault} (SAF), \textit{transition fault} (TF), \textit{coupling fault} (CF). Based on the number of involved cells,  memory cell array faults can be further classified into: \textit{single cell faults} (\#C$=$1), \textit{coupling faults} (\#C$=$2), and \textit{neighborhood pattern sensitive faults} (\#C$>$2). Read\&write circuit faults are those speed-related faults occurring in the read\&write circuits, such as \textit{slow sense amplifier fault} (SSAF) and \textit{slow write driver fault} (SWDF). As  memory cell arrays occupy the majority part of a memory chip, research efforts mainly focus on faults in this location while faults in the other two locations are somewhat neglected. This is also the case for STT-MRAMs. All existing publications are dedicated to the analysis and detection of faults in memory cell arrays whereas faults in other locations have not been explored so far, to the best of our knowledge.

\subsubsection{Lifetime}
 Based on the fault lifetime, faults can also be classified into two categories: permanent faults and transient faults \cite{Zhao2016, Kang2015a, Kang2015}. The former refer to those faults that are permanent and uncorrectable, which can be caused by manufacturing defects and extreme process, voltage, or temperature (PVT) variations. Permanent faults feature cell-to-cell variation, which means that only a small fraction of memory cells are defective and thus do not function as expected, while the majority of cells function as well as intended in designs. Due to the deterministic effects of defects and PVT variations on memory cells after fabrication, permanent faults are  generally fixed and detectable with with certain testing techniques. The latter are those faults that are temporary, and they can be self-corrected by a new operation on the faulty cell. Instead of the cell-to-cell variation for permanent faults, transient faults introduce cycle-to-cycle variation. In other words, the fault occurrences are probabilistic and unpredictable during the lifetime of memory chips, due to the STT stochastic switching nature and thermal noise, etc. Therefore, transient faults should not be considered as the target during manufacturing tests \cite{Nair2018a}. However, it is worth noting that  transient faults are  increasingly becoming a reliability challenge, since PVT variations, thermal fluctuation, magnetic coupling, and radiation   are no longer neglectable interference factors as technology scales down. Consequently, stronger ECCs or other correction techniques are required to correct these run-time faults.

\subsubsection{Physical nature}
Last but not least, faults can also be grouped into static faults and dynamic faults based on their physical natures. The former are generally not timing-related and require at most one operation (\#O$<=$1) to be sensitized. The latter are often timing-related and require more than one consecutive operations (\#O$>$1) to be sensitized.

\begin{figure}[!t]
	\centering
	\includegraphics[width=0.5 \textwidth ]{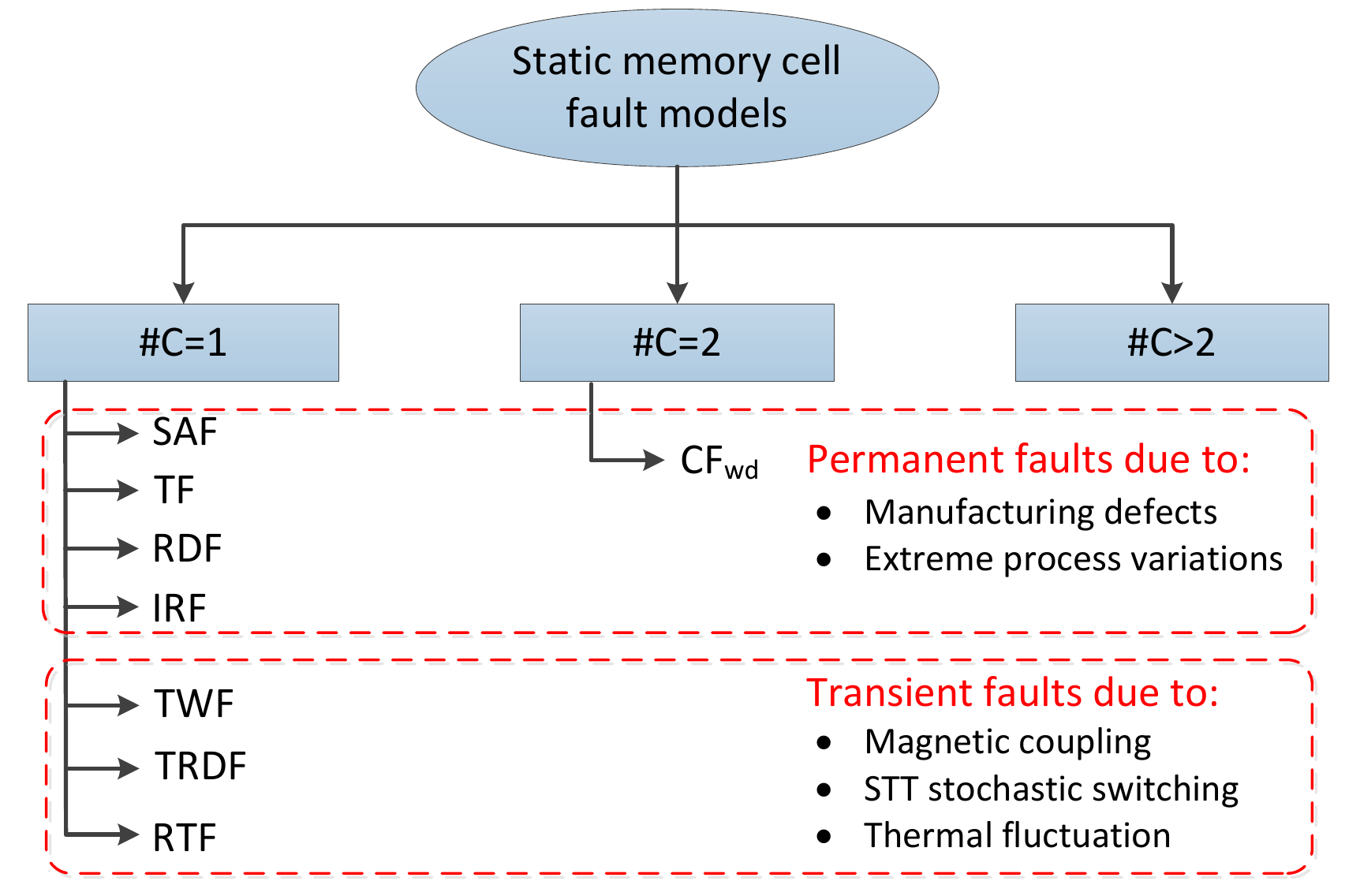}
	\caption{Classification of static faults in  STT-MRAM cell array.}
	\label{fig_memcell_classification}
\end{figure}

\subsection{Static Faults in STT-MRAM Cell Arrays}
\label{subsec:static_faults}
In this part, we will mainly examine  static faults occurring in STT-MRAM cell arrays, since research attempts to date are mainly of this category due to the dominant area of cell arrays in STT-MRAM chips. Fig. \ref{fig_memcell_classification} shows the classification of all fault models  that have been proposed in the literature in this category. Next,  we will first introduce permanent fault models, including stuck-at-fault (SAF), transition fault (TF), read destructive fault (RDF), incorrect read fault (IRF), and write disturb coupling fault (CF\textsubscript{wd}). Thereafter, we will elaborate  transient faults including transient write fault (TWF), transient read disturb fault (TRDF), and retention fault (RTF), which are mainly caused by STT-MRAM specific failure mechanisms such as STT stochastic switching nature and thermal perturbation.
\begin{figure}[!t]
	\centering
	\includegraphics[width=0.43 \textwidth ]{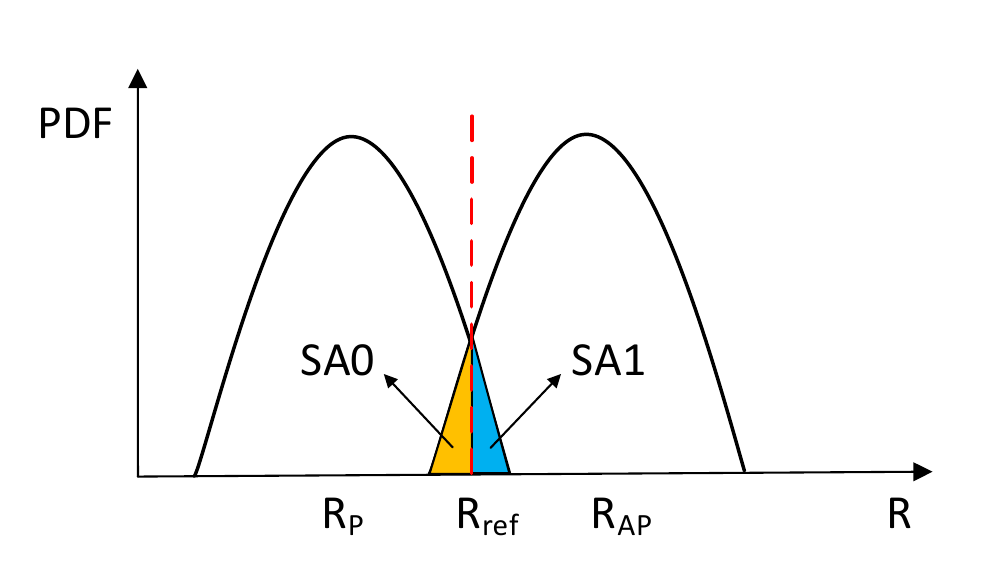}
	\caption{Illustration of the SAF caused by resistance distribution tail overlap of $R_{\mathrm{P}}$ and $R_{\mathrm{AP}}$ due to extreme process variation in  $t_{\mathrm{ox}}$ of the MTJ device.  }
	\label{fig_SAF}
\end{figure}
\subsubsection{Permanent Faults}

\emph{\textbf{Stuck at fault (SAF)}}: this fault refers to that a cell always presents logical value 0 (SA0) or 1 (SA1), no matter what values are written into it. SA0 and SA1 faults are denoted as $<$$\forall/0/-$$>$ and $<$$\forall/1/-$$>$, respectively. Similar to  traditional charge-based RAMs, SAFs in STT-MRAMs can be caused by physical defects. For instance, the authors in \cite{Chintaluri2016a} proposed that some resistive shorts or bridges (i.e., electrical equivalents to certain physical defects) in the STT-MRAM cell array can cause SAFs. Furthermore, process variations tend to deviate  key MTJ and transistor parameters  from their nominal values, leading to SAFs under extreme circumstances \cite{Kang2015a}. For example, the author in \cite{Zhao2016} observed that the MTJ tunnel barrier exhibits different thickness arranging from \SI{0.86}{\nano\meter} to \SI{1.07}{\nano\meter}, while the nominal value is \SI{1}{\nano\meter}. As the MTJ resistance is exponentially dependent on the tunnel barrier thickness $t_{\mathrm{ox}}$, a tiny $t_{\mathrm{ox}}$ variation will lead to a huge difference in resistance. Fig. \ref{fig_SAF} illustrates that the $t_{\mathrm{ox}}$ variation causes a partial overlap of  $R_{\mathrm{P}}$ and $R_{\mathrm{AP}}$ distributions.  An MTJ device with $R_{\mathrm{P}}$ falling in the tail over $R_{\mathrm{ref}}$  would suffer from a SAF1 fault, whereas the lower tail of $R_{\mathrm{AP}}$ distribution under $R_{\mathrm{ref}}$ would lead to a SA0 fault.  As technology scales down, reducing process variation in $t_{\mathrm{ox}}$ becomes increasingly challenging during fabrication.

\begin{figure}[!t]
	\centering
	\includegraphics[width=0.35 \textwidth ]{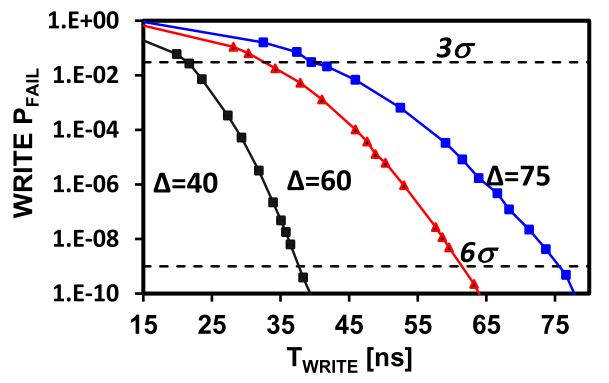}
	\caption{Write error rate ($P_{\mathrm{FAIL}}$) as a function of the required write time ($T_{\mathrm{WRITE}}$) with $\Delta=40, 60, 75$ . The horizontal dashed line labeled as $6\sigma$ represents the $6\sigma$-corner requirement for the write time of a write operation with  $P_{\mathrm{FAIL}}=10^{-9}$. Reprinted from \cite{Chintaluri2016a}. }
	\label{fig_writeFailureProbability}
\end{figure}

\begin{figure}[!t]
	\centering
	\includegraphics[width=0.35 \textwidth ]{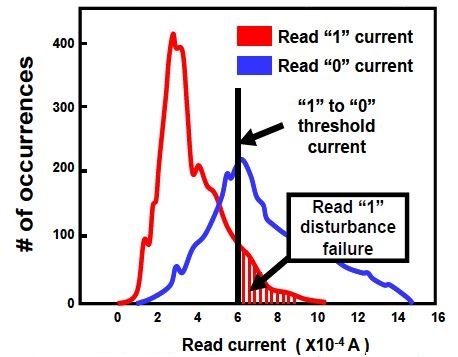}
	\caption{Read current distribution caused by process variations. The read ``1'' currents for some worst-case cells exceeding the 1$\rightarrow$0 threshold current lead to RDFs. Reprinted from \cite{Roy2014s}.}
	\label{fig_P_readdisturb}
\end{figure}

\begin{figure*}[!t]
    \centering
    \subfloat[]{\includegraphics[width=0.43 \textwidth]{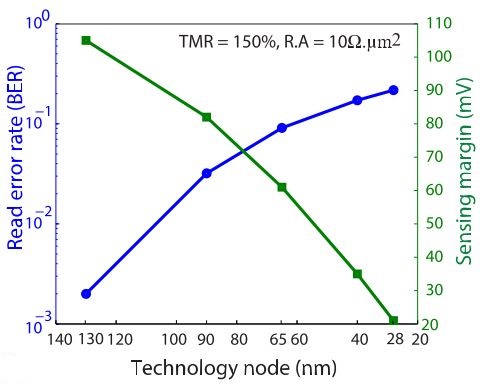}%
    \label{fig_TechnologyShrinking}}
    \hfil
    \subfloat[]{\includegraphics[width=0.41 \textwidth]{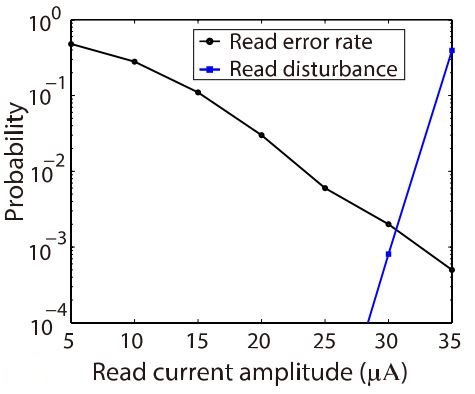}%
    \label{fig_ReadCurrentShrinking}}
    \caption{(a) Read error rate (i.e., IRF) and sensing margin vs. technology scaling. (b) Read error rate and read disturbance (i.e., RDF) vs. read current amplitude at 28nm technology node. Reprinted from \cite{Kang2015}.}
    	
    \label{fig_IRF}
\end{figure*}

\emph{\textbf{Transition Fault (TF)}}: this fault refers to that a cell fails to make a rising transition (0$\rightarrow$1) or a falling transition (1$\rightarrow$0) when it is written. TF includes TF0 and TF1, which are  denoted  as $<$$0\mathrm{w}1/0/-$$>$ and $<$$1\mathrm{w}0/1/-$$>$, respectively. TFs can be caused by defects as well as extreme process variations. For example, resistive opens along the write current path in Fig. \ref{fig_basicOps}(a) lead to a current degradation \cite{Chintaluri2016a}. As a result, TFs are likely to take place due to a lack of enough current passing through the MTJ device ($I_{\mathrm{w}}$$<$$I_\mathrm{c}$). Additionally, resistive opens in the word line  have an effect of limiting the current driving ability of the access NMOS in  bit-cells, thereby leading to TFs. 

 Process variations may also result in TFs in some cases. As we discussed previously, in order to switch successfully between AP and P states, a write operation has to supply both sufficient  amplitude and duration of the write current to overcome the energy barrier between them. In this sense, PV-induced variation in the required switching time ($t_{\mathrm{sw}}$) for individual bit-cells may make some cells  subject to TFs since the STT-switching action does not finish within the fixed write time in practical circuit designs ($ t_{\mathrm{sw}}$$>$$t_{\mathrm{wr}}$). E. Vatajelu et al. referred to this PV-induced TF as \textit{slow write fault} (SWF) \cite{Vatajelu2017}. Chintaluri et al.[xxx] observed that write operations are more sensitive to $V_{\mathrm{th}}$, $t_\mathrm{{ox}}$, and $M_{\mathrm{s}}$ than  other parameters of the MTJ and transistor.  Fig. \ref{fig_writeFailureProbability} shows the write error rate of a write operation as a function of the write time for MTJ devices with some given thermal stability values. It can be seen that memory cells with $\Delta=40$ require a write time of at least $\sim$\SI{40}{\nano\second}  to reach a write error rate $P_{\mathrm{FAIL}}=10^{-9}$ (equivalent to  $6\sigma$-corner requirement). In other words, TFs happen to those cells that require more than \SI{40}{\nano\second} to successfully switch to the other state. 
 As the thermal stability increases, indicating longer retention time, it requires longer write time to maintain the same write error rate.  It is worth noting that the continuous technology down-scaling and adoption of PMA-MTJ devices, on one hand, result in a dramatic reduction of the  MTJ critical switching current density ($J_{\mathrm{c0}}$), which greatly alleviates the TF issue. On the other hand, the probability of read disturbance rises because the gap between the read current  $I_{\mathrm{rd}}$ and the critical switching current  $I_{\mathrm{c}}$ narrows considerably \cite{kim2011extended}.

\emph{\textbf{Read Destructive Fault (RDF)}}: this fault arises when a read operation causes an inadvertent flip in the addressed bit-cell. For STT-MRAMs, only RDF1 (denoted as $<$$1\mathrm{r}1/0/0$$>$) is possible to occur due to the uni-directional property of read operations, as shown in Fig. \ref{fig_basicOps}(d). As $I_{\mathrm{w0}}$ and $I_{\mathrm{rd}}$ share the same path in the 1T-1MTJ bit-cell structure, a read current is possible to behave as a weak write current and therefore results in an unintended magnetization flip in the FL in the presence of some resistive defects\cite{Chintaluri2016a}. Apart from  defects, process variations are also likely to cause a RDF1. For instance, a low $R_{\mathrm{AP}}$ of the MTJ, a low $V_{\mathrm{th}}$ of the  NMOS, or a degraded $\Delta$ elevate the read current $I_{\mathrm{rd}}$ above the critical switching current $I_{\mathrm{c}}$ for some worst-case cells, as shown in Fig. \ref{fig_P_readdisturb} \cite{Raychowdhury2009,Augustine2011,Fong2014}. It is worth noting that RDF0 ($<$$0\mathrm{r}0/1/1$$>$) can never happen when the cell is in the P state, since a P$\rightarrow$AP transition requires a   current flowing from  SL to BL whereas a read operation invokes a reversed current. 

\emph{\textbf{Incorrect Read Fault (IRF)}}: this fault refers to a sensing failure of  the actual resistive state  in the addressed cell. IRF includes IRF0 and IRF1, which are denoted as  $<$$0\mathrm{r}0/0/1$$>$ and $<$$1\mathrm{r}1/1/0$$>$, respectively. Resistive opens along the read current path can lead to a  decrease in the read current. This causes a read operation addressed at a  cell with the above defects in the P state  to return an incorrect logical value ``1'' ($<$$0\mathrm{r}0/0/1$$>$) \cite{Chintaluri2016a}. In addition, resistive bridges shorting  the SL and the internal node of the memory cell can pull up the read current, therefore leading to a IRF1 ($<$$1\mathrm{r}1/1/0$$>$) when the addressed cell is in the AP state.

 Process variations also contribute to IRFs. In order to correctly read the resistive state of the MTJ (0 for $R_{\mathrm{P}}$ and 1 for $R_{\mathrm{AP}}$), a minimal sensing margin  is required irrespective of sense amplifier designs. the sensing margin is defined as the gap between the current going through the cell under sensing and the current going through the reference cell. In this regard, a high TMR and less process variations are crucial to guarantee a large sensing margin. However, as technology scales down, the read reliability is increasingly challenging and becoming a bottleneck in STT-MRAM circuit designs. Fig. \ref{fig_IRF}\subref{fig_TechnologyShrinking} shows the climbing trend of read error rate ( i.e., IRF) and the decrease in the  as the technology node becomes smaller, caused by deterioration of process variation and TMR parameter \cite{Kang2015a}.  Furthermore, the conflict between read error rate and read disturbance in optimizing read current to achieve better readability, shown in Fig. \ref{fig_IRF}\subref{fig_ReadCurrentShrinking}, again squeezes the design space for reliable read operations \cite{Kang2015}.

\emph{\textbf{Write Disturb Coupling Fault (CFwd)}}: this fault arises when a write operation on a bit-cell (aggressor) results in an unintended flip of another bit-cell (victim). Notations for CFwd are $<$$x\mathrm{w}$$\sim$$x; 0/1/-$$>$ or $<$$x\mathrm{w}$$\sim$$x; 1/0/-$$>$ for a transition write  on the aggressor cell, and $<$$x\mathrm{w}x; 0/1/-$$>$ or $<$$x\mathrm{w}x; 1/0/-$$>$ for a  non-transition write  on the aggressor cell. For example, $<$$x\mathrm{w}$$\sim$$x; 0/1/-$$>$ means a write operation on the aggressor cell to flip its state from $x$ to $\sim$$x$ inadvertently makes the victim cell flip from 0 to 1.  Typical causes of CFwd include  the stuck-at-ON defect of the access transistor in the bit-cell, causing an extra write to the victim cell when writing adjacent cells sharing the same BL and SL. Inter-cell bridges between WLs are also reported as the cause of  CFwd \cite{Chintaluri2016a}. It is worth noting that researchers from different institutes use different terminologies to describe CFwd. Chintaluri et al. from  Georgia Institute of Technology refer to this fault as a coupling fault \cite{Chintaluri2016a, Insik_ITC_2016}, whereas Vatajelu et al. from TIMA Laboratory (France) call it a write disturb fault \cite{Vatajelu2017}. Combining both of them, we believe that the term write disturb coupling fault is more appropriate  based on its faulty behavior and causes.

\subsubsection{Transient Faults}
We have  discussed  permanent faults caused by defects and extreme process variations so far. Those faults also exist in conventional memory technologies such as SRAMs and DRAMs. As STT-MRAM is an emerging NVM technology based on many novel physical phenomena, transient faults which are intermittent in some cycles have increasingly been a reliability concern. In this part, we will start with introducing all influential factors that impact the reliability of STT-MRAMs. Thereafter, some resulted transient faults will be discussed.

\emph{\textbf{Transient Write Fault (TWF)}}: Due to the stochastic property of STT-switching behavior,  the incubation time before the actual start of magnetization precession in the FL  varies from one event to the next, as  mentioned previously. However, the current pulse width of write operations is generally fixed with some margin over the averaged switching time for STT-MRAM circuit designs in practice. This  may lead to an unexpected write fault when the incubation delay is longer than the given pulse width of  write currents. We refer to this fault as \textit{transient write fault} (TWF) in this paper. TWF includes TWF0 and TWF1, which are denoted as $<$$0\mathrm{w}1/T0/-$$>$ and $<$$1\mathrm{w}0/T1/-$$>$, respectively.

The major difference between TWF and previous TF is that the former is intrinsically unpredictable and temporary while the latter is deterministic and permanent. Specifically, TWF may happen to all STT-MRAM  cells, including those defect-free cells, with a very small probability. It can be self-repaired by the next write operation directly following a faulty one. TF, however, occurs at the cells with some defects or extreme process variations aforementioned, and it always leaves a faulty state in those cells after a transition write operation.

TWF is aggravated by thermal fluctuation, current asymmetry in w0 and w1 operations, and PVT variations \cite{Zhang2011, Jones2012}. First, due to thermal fluctuation, the actual switching time has a wide distribution over cycles; the effects of thermal fluctuation on the STT-induced switching process can be modeled by introducing a thermally-induced random field $\bm H_{\mathrm{fluc}}$ and an initial angle $\theta$ between the magnetizations in the FL and the PL (see Equations (\ref{eq:LLG_ThermalFluc})). Second, the current asymmetry in w0 and w1 operations ($I_{\mathrm{w0}}$$>$$I_{\mathrm{w1}}$), caused by the  source degeneration of the access NMOS \cite{Lee2012, Jones2012}, further widens the  switching time distribution. Third, the increasing PVT variations with technology downscaling also induce a wide switching time distribution over STT-MRAM cells. Considering above three factors, a large write margin (i.e., a long pulse to cover the wide distribution of the switching time) is required to guarantee a high switching probability for all cells and cycles. However, this comes with the cost of sacrificing write performance and energy.   Hence, TWF is increasingly  posing a  threat to the reliability of STT-MRAM designs.

\begin{figure}[!t]
	\centering
	\includegraphics[width=0.43 \textwidth ]{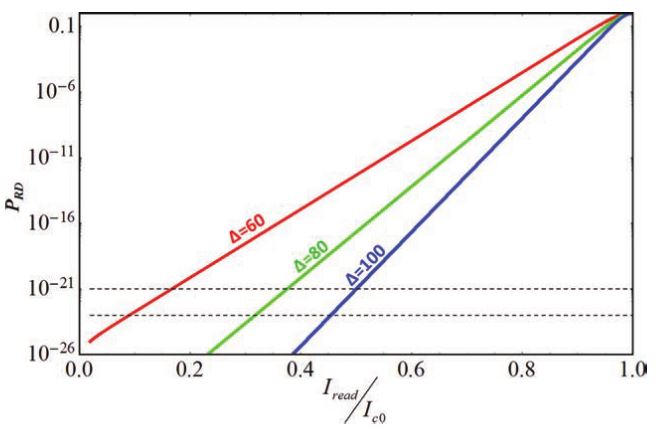}
	\caption{TRDF probability as a function of read current ($\SI{10}{\nano\second}$) with respect to various $\Delta$ values. Typical requirement of TRDF probability is in the range of $10^{-23}$ to $10^{-21}$ (the two horizontal dotted lines), setting up a ceiling to the read current value for STT-MRAM cells with various $\Delta$ values. Reprinted from \cite{Apalkov2013}. }
	\label{fig_TRDF}
\end{figure}

\emph{\textbf{Transient Read Disturb Fault (TRDF)}}: Due to thermal fluctuation, the state of a cell may accidentally flip during a read operation despite the read current is  much smaller than the critical switching current \cite{Bishnoi2015, Ran2016}. We name this fault as  \textit{transient read disturb fault} (TRDF). Similar to the aforementioned RDF,  only TRDF1 (denoted as $<$$1\mathrm{r}1/T0/1$$>$) exists in STT-MRAMs since the read current is uni-directional. Nevertheless, TRDF1 is essentially different from RDF1 which is caused by some defects or extreme process variations, as we discussed previously. First, TRDF1 happens stochastically to all bit-cells, rather than to some cells with defects or extreme PVs, in STT-MRAM chips with a very small probability. The probability of TRDF1 can be approximately calculated by the Neel-Brown model (see Equation (\ref{eq:readdistrub})). Second, TRDF1 is inherently caused by thermal fluctuation, which is strengthened by applying a read current to the MTJ device due to Joule heating, whereas RDF1 is the result of overdriven read currents, due to defects or extreme PVs. Third, TRDF1 is not reproducible, which means the majority of read 1 operations success without any destruction to the accessed cell while a very small fraction of them end up flipping the cell from 0 to 1 state. By contrast,  RDF1 is reproducible meaning that all read 0 or 1 operations lead to a state flip in the target cell. Despite above differences between TRDF1 and RDF1, the prerequisite for them are same: $I_{\mathrm{w0}}$ and $I_{\mathrm{rd}}$ share the same path and the target MTJ device must be in AP state. We give a brief comparison between RDF1 and TRDF1 in Table \ref{table:RDF_TRDF}.

TRDF in STT-MRAMs can be alleviated by enlarging the gap between the read current and the critical switching current.  This can be achieved by  either increasing the critical switching current or reducing the read current. Apparently, increasing the critical switching current is not practical because the high write current and power dissipation are already a bottleneck in STT-MRAM designs.  In reality, the critical switching current is continually going down as the size of MTJ device shrinks \cite{Ran2016}.  On the other hand, reducing the read current  is a possible solution. But it also squeezes the read margin, posing a threat to the read  reliability for sense amplifier designs. Based on Equation (\ref{eq:readdistrub}), one can observe that  a shorter read pulse translates to smaller read disturb probability. This can also be considered in practical circuit design while maintaining the effectiveness of sense amplifier voltage development.  Despite these possible solutions to reduce the probability of TRDF, it is still going to be a major reliability issue in future technology nodes, as pointed out in \cite{Bishnoi2015}. Fig. \ref{fig_TRDF} shows the TRDF probability as a function of read current with a pulse width of  \SI{10}{\nano\second} with respect to various $\Delta$ values. For a typical requirement of TRDF probability from $10^{-23}$ to $10^{-21}$,  the maximum allowed read current is around $0.1I_{c0}$ for  STT-MRAM cells with $\Delta=60$, posing a big challenge to the sense amplifier design in order to obtain reliable and low-latency read operations.
\begin{figure}[!t]
	\centering
	\includegraphics[width=0.42 \textwidth ]{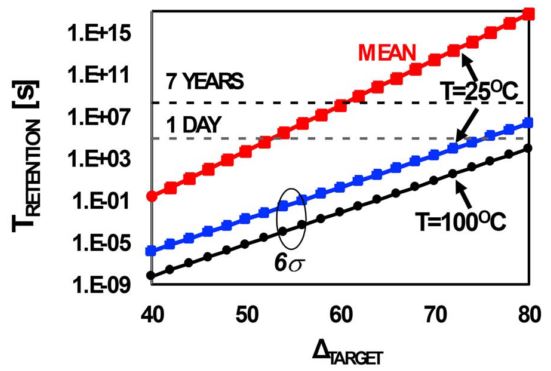}
	\caption{Retention time vs. target $\Delta$ value considering process variation (mean and $6\sigma$ corner) and temperature (\SI{25}{\celsius} and \SI{100}{\celsius}) Reprinted from \cite{Chintaluri2016a}. }
	\label{fig_Delta_PV}
\end{figure}

\begin{table}[!t]
  \centering
  \caption{A  comparison between RDF1 and TRDF1. }
  \resizebox{\columnwidth}{!}{%

    \begin{tabular}[m]{|c|c|c|}

    \hline
    Fault Models & RDF1  & TRDF1        \\

    \hline
      Prerequisite             &  \multicolumn{2}{c|}{  \makecell[l]{       \:\:\:\:\:\:\:\:\:\:\:\:\:\:\:\:\:\:\:1. $I_{\mathrm{w0}}$ and $I_{\mathrm{rd}}$ share the same path \\ \:\:\:\:\:\:\:\:\:\:\:\:\:\:\:\:\:\:\:2. MTJ in AP state } }  \\

    \hline
     Victim Cells               &  Cells with defects or extreme PVs  &  All cells  \\
         \hline
     Causes                     &  $I_{\mathrm{rd}}>I_{\mathrm{c}}$                    &  Thermal fluctuation  \\
         \hline
     Occurrence Probability     &  Absolute                           &  Small probability event \\
         \hline
     Repeatability              &  Yes                                &  No                    \\
         \hline
    Readout data                & Wrong                              & Correct   \\
        \hline
    Notation                    & $<1\mathrm{r}1/0/0> $                          & $<1\mathrm{r}1/T0/1>$  \\
        \hline
    \end{tabular}}
  \label{table:RDF_TRDF}
\end{table}
\emph{\textbf{Retention Fault (RTF)}}: This fault means a cell loses it content over time, due to thermal fluctuation. RTF includes RTF0 and RTF1 which are denoted as  $<$$0_T/1/-$$> $ and $<$$1_T/0/-$$>$, respectively. As a static model, Equations (\ref{eq:E_B}-\ref{eq:retention}) give an approximation of  the retention time for STT-MRAMs. It suggests that RTF exponentially depends on the MTJ dimension and the ambient temperature, but it does not imply the cause of RTF in STT-MRAMs. Unlike the retention fault in DRAMs where the amount of charge on cell capacitors decreases gradually, retention fault in STT-MRAMs take place instantly (a stochastic process)  in the presence of thermal noise \cite{naeimi2013sttram}. Accordingly, the retention time of every cell in STT-MRAMs is not fixed and predictable in essence. Rather, it fluctuates dynamically depending on the intensity of thermal perturbation. Intel Technology Journal \cite{naeimi2013sttram} suggests that a bit flip induced by thermal fluctuation has a Poisson distribution with a time characteristic $\tau_0 e^{\Delta}$. Thus, the retention fault happens when the bit-cell flips with an odd number of time. If we set $I=0$ in Equation (\ref{eq:readdistrub}), $\tau_1$ now becomes the nominal retention time, and the probability of RTF can be calculated by:

\begin{equation}
  P_{RTF}=1-\exp(-\frac {t}{\tau_0\exp(\Delta)}),    \label{eq:retention_failure}
\end{equation}
where $t$ is the observation time window, and $\tau_0$ is the attempt time ($\sim$\SI{1}{\nano\second}).

Furthermore, in the case of $I\neq0$ (e.g., during read  operations), the thermal fluctuation is strengthened, thus raising the probability of magnetization reversal. Besides, the value of $\tau_1$ decreases significantly due to the degraded thermal stability $\Delta_I$ under a read or write current:
\begin{equation}
  \Delta_I=\Delta_0(1-\frac {I} {I_{c0}}).    \label{eq:Delta_I}
\end{equation}
This phenomenon can be utilized to experimentally measure the norminal thermal stability (i.e., $\Delta_0$ in Equation (\ref{eq:Delta_I})) of MTJ devices.  By repeatedly applying a weak current through the target MTJ device,  a statistical probability of magnetization reversal can be obtained to calculate $\Delta_I$ \cite{Heindl2011}. Thereafter, $\Delta_0$ can be calculated based on Equation (\ref{eq:Delta_I}).

\begin{figure}[!t]
	\centering
	\includegraphics[width=0.4 \textwidth ]{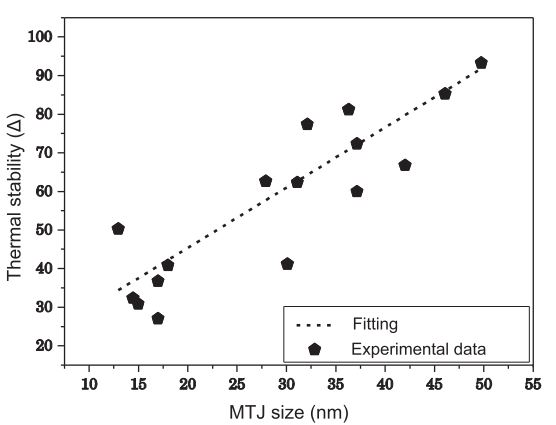}
	\caption{Experimental measurement and fitting curve of the thermal stability $\Delta$ with various MTJ sizes. Reprinted from  \cite{Ran2016}. }
	\label{fig_Delta_TS}
\end{figure}

Since retention time (i.e., the average lifetime, approximated by Equation (\ref{eq:retention})) of STT-MRAM cells is exponentially dependent on thermal stability, any factors that degrade $\Delta$ would reduce the  retention time. Obviously, the smaller the $\Delta$ value, the more vulnerable to RTF. First, an increase in temperature would exponentially boost thermal fluctuation and  degrade the $\Delta$ value,  thereby  reducing the average retention time as shown in Fig. \ref{fig_TMR_Reten_T}\subref{fig_retentionT}.
Second, process variations inevitably introduce $\Delta$ variation among  STT-MRAM cells. Fig. \ref{fig_Delta_PV} shows the  retention time as a function of the target $\Delta$ for mean cells and $6\sigma$-corner cells. For STT-MRAM cells with a target $\Delta$ of 60, the retention time of mean cells is around 7 years under room temperature, but the $6\sigma$-corner cells (lower $\Delta$) have a retention time far less than 1 day. If the temperature rises to \SI{100}{\celsius}, the retention time becomes even shorter.  Third, magnetic coupling is another concern that impacts the thermal stability of memory cells depending on the data pattern in the neighborhood. Fig.~\ref{fig_Delta_MC} shows that the worst data pattern generates a stray field that significantly degrades the $\Delta$ value at various technology nodes. Fourth, a current flowing through the MTJ device during write and read operations degrades the $\Delta$ value as well, as seen in Equation (\ref{eq:Delta_I}). Last but not least, thermal stability decreases as the size of MTJ devices shrinks, as shown in Fig. \ref{fig_Delta_TS}. This is because a smaller FL of the MTJ  leads to a lower energy barrier, depicted by Equation (\ref{eq:E_B}). Experimental results in \cite{Iyengar2016} show that a 2x shrinkage in the MTJ volume decreases the retention time from a few decades to a few seconds.  To obtain the required $\Delta$ value at advanced technology nodes, the anisotropy field $H_{\mathrm{k}}$ in the FL of the MTJ must be enhanced, which makes the manufacture of STT-MRAMs become more and more challenging as technology scales down. Due to those influential factors that may degrade the thermal stability of STT-MRAMs, RTF is predicted to be the dominating threat to STT-MRAM reliability at future technology nodes \cite{naeimi2013sttram}.

\pdfoutput=1

\section{TEST ALGORITHMS }
\label{sec:test_algorithms}
In this section, we will first discuss March test algorithms proposed in the literature. These March tests can guarantee the detection of  certain strong functional fault models such as SAF and TF. Thereafter, we will introduce and compare three test algorithms dedicated to testing the  retention time of STT-MRAMs.

\subsection{March Tests}
\label{subsec:MarchTest}
March tests are commonly used in detecting traditional memory faults, such as SAF, TF, etc., due to their linear complexity, regularity, and symmetry \cite{Bushnell2000a}.  A March test consists of a finite sequence of march elements, each of which is composed of a sequence of read and/or write operations applied to each memory cell before proceeding to the next. In the literature, there are several papers related to testing MRAMs with March tests.   

Chin et al.  presented in \cite{Su2004,Chien2006} a fault model called write disturb fault (WDF) in toggle MRAMs, where the magnetic field generated in write operations to switch the state of the addressed bit-cell (aggressor) may inadvertently reverse the data stored in adjacent cells (victims). Furthermore, they proposed  March C- and March 17N test algorithms to detect SAF, TF, CF, AF, and WDF in their demonstration toggle MRAM chip.

In STT-MRAMs, the cause of WDF is totally different from traditional toggle MRAMs. Rather than induced by writing magnetic field of current-carrying wires in toggle MRAMs, WDF in STT-MRAMs is caused by defects such as access transistor stuck-at-ON or resistive bridges between cells. In this paper, we refer to WDF in STT-MRAMs as write disturb coupling fault (CFwd) as discussed in Section \ref{subsec:static_faults}(1), since it involves more than 1 bit-cell.  To detect CFwd as well as SAF, TF, RDF, and IRF in STT-MRAMs, Yoon et al. proposed in \cite{Insik_ITC_2016} a word-oriented March (WOM) test algorithm (26N) as follows.
\begin{align*}
  &  \Updownarrow (\mathrm{w}00); \Uparrow (\mathrm{r}00,\mathrm{w}11,\mathrm{r}11);\Downarrow (\mathrm{r}11,\mathrm{w}00,\mathrm{r}00); \\
  & \Downarrow (\mathrm{r}00,\mathrm{w}11,\mathrm{r}11);\Downarrow (\mathrm{r}11,\mathrm{w}00); \Updownarrow (\mathrm{r}00); \\
  & \Uparrow (\mathrm{r}00,\mathrm{w}01,\mathrm{r}01);\Uparrow (\mathrm{r}01,\mathrm{w}10,\mathrm{r}10);\Downarrow (\mathrm{r}01,\mathrm{w}10,\mathrm{r}10);\\
  & \Downarrow (\mathrm{r}10,\mathrm{w}11,r11);\Updownarrow (\mathrm{r}11)\\
\end{align*}
Recently in ITC2018, Nair et al. reported  \textit{dynamic incorrect read fault} (dIRF)  based on their circuit simulations. dIRF is an incorrect read fault which is sensitized by at least two consecutive read operations due to a resistive bridge defect between the SL and the internal node of the 1T-1MTJ cell. To detect dIRF, they proposed the following March algorithm (14N).
\begin{align*}
&  \Uparrow (\mathrm{w}0);  \\
& \Uparrow (\mathrm{r}0, \mathrm{w}1, \mathrm{r}1, \mathrm{r}1, \mathrm{r}1, \mathrm{r}1); \\
& \Downarrow (\mathrm{r}1, \mathrm{w}0, \mathrm{r}0, \mathrm{r}0, \mathrm{r}0, \mathrm{r}0);\\
& \Downarrow (\mathrm{r}0)\\
\end{align*}

\subsection{Retention Time Tests}
\label{subsec:retentiontest}
In Section \ref{subsec:static_faults}(2), we have discussed the characteristics, causes, and influencing factors of retention fault in STT-MRAMs. Yet, the requirement of retention time for STT-MRAMs ranges from a few seconds to ten years, depending on the specific application. For the storage-class memory (SCM) application, $10+$ years of retention time is desired as most of the data is rarely revisited. This requires a thermal stability ($\Delta$) as high as $80$, which is experimentally achievable at the expense of high write energy and latency.  In contrast, for the last-level-cache application which is seen as one of the most important markets for applying STT-MRAM technology in the short term, the average revival time of cache blocks is below \SI{1}{\second} \cite{Jog2012a}. This allows us to  trade the retention time of STT-MRAMs for better write performance, which has been  extensively studied \cite{smullen2011relaxing,li2013low,sun2014stt}.

Regardless of the different requirements of  retention time of STT-MRAMs for different applications, testing retention time of STT-MRAMs is very important. However, characterizing  STT-MRAM retention time is very challenging, since retention fault is essentially a transient and stochastic fault which depends on temperature, process variations, magnetic perturbation, and disturb current.  Thus,  March tests  are not suitable to test it, and traditional retention tests for DRAMs cannot be directly applied to STT-MRAMs.  This calls for special test techniques. Next, we will introduce three test approaches dedicated to testing STT-MRAM retention time.  
 \begin{algorithm}[!t]
	\caption{Retention time based on weak disturb current.}
	\begin{algorithmic}
		\renewcommand{\algorithmicrequire}{\textbf{Input:}}
		\renewcommand{\algorithmicensure}{\textbf{Output:}}
		\REQUIRE $I_{\mathrm{wwr}}[N]$ = array containing N number of $I_{\mathrm{wwr}}$ values
		\REQUIRE $t$ = the current pulse width of $I_{\mathrm{wwr}}$
		\REQUIRE $M$ = the number of experiments for each $I_{\mathrm{wwr}}$ value
		\ENSURE  Retention time for a cell
		\STATE Initialization
		\FOR {$i=0$ to $N-1$}
		\STATE Regular write of a test pattern
		\FOR {$j=0$ to $M-1$}
		\STATE Weak write with current $I_{\mathrm{wwr}}[i]$ for time $t$
		\STATE Regular read
		\IF {$readout \: data \ne test \: pattern$}
		\STATE Error counter ++
		\STATE Rewrite the test pattern
		\ENDIF
		\ENDFOR
		\STATE Pr[i]=Error counter/M
		\STATE Reset error counter
		\ENDFOR
		\STATE Extrapolation of $\Delta$ with equation (\ref{eq:TestDelta})
		\STATE Approximation of $T_{\mathrm{ret}}$ with equation (\ref{eq:retention})
		\RETURN $T_{\mathrm{ret}}$
	\end{algorithmic}
\end{algorithm}
\subsubsection{Statistic method With a Weak Disturb Current}
Intel proposed a method to test the retention time of STT-MRAM cells by applying a weak disturb current through them \cite{naeimi2013sttram}. Equation (\ref{eq:readdistrub}) can be used to calculate the switching probability in the thermal activation regime under a long ($t_{\mathrm{p}}$) but weak write current ($I_{\mathrm{wwr}}<I_{\mathrm{c}0}$). In this case, we derive:
\begin{equation}\label{eq:estimation}
  \frac{t_{\mathrm{p}}}{\tau_0 \exp(\Delta(1-I_{\mathrm{wwr}}/I_{\mathrm{c}0}))}<<1.
\end{equation}
By performing Taylor expansion, we can derive the following expression:
\begin{equation}\label{eq:TestDelta}
  \ln(Pr(I_{\mathrm{wwr}}))=\ln(\frac{t_{\mathrm{p}}}{\tau_0})-\Delta(1-\frac {I_{\mathrm{wwr}}} {I_{\mathrm{c}0}}).
\end{equation}
Equation (\ref{eq:TestDelta}) links the \textit{disturb probability} $Pr(I_{\mathrm{wwr}})$ with the thermal stability $\Delta$ under a  long  but weak write current. This equation can be used to experimentally measure the $\Delta$ value.  By repeatedly applying  a large number of  weak write currents (e.g., $I_{\mathrm{wwr}}<0.8I_{\mathrm{c}0}, t=\SI{100}{\nano\second}$) to the memory cell under test, we can obtain a  statistic result for $Pr(I_{\mathrm{wwr}})$. Thereafter,  $\Delta$ value can be derived with Equation (\ref{eq:TestDelta}). As the retention time is exponentially dependent on $\Delta$, it can thus be approximated according to Equation (\ref{eq:retention}). The complete test process is described with Algorithm 1.

Though theoretically feasible, the test time using this method  is prohibitive in practice. In order to get a statistic result for the $Pr(I_{\mathrm{wwr}})$ with a $1\%$ error margin, $5\times10^5$ number of tests are needed for each data point assuming that the expected probability  is $1\times10^{-3}$. Consequently, it takes approximately \SI{0.5}{\second} to test a bit-cell and more than 5 days to test a 64-MB array \cite{naeimi2013sttram}. Note that this is only the time spent on estimating the $Pr(I_{\mathrm{wwr}})$ for all cells in the array without taking into account the subsequent calculation of $\Delta$ and retention time. Worse still, the test time increases with the array size and $\Delta$ value. Hence,  this is obviously unacceptable from the perspective of test time.

\begin{algorithm}[!t]
	\caption{Burn-in retention test based on binary search.}
	\begin{algorithmic}
		\renewcommand{\algorithmicrequire}{\textbf{Input:}}
		\renewcommand{\algorithmicensure}{\textbf{Output:}}
		\REQUIRE $N_{\mathrm{seh}}$ = the number of binary searches for averaging
		\REQUIRE $N$ = iterations in a search
		\REQUIRE $t_{\mathrm{UB}}$ = upper bound of the predicted retention time
		\REQUIRE $t_{\mathrm{LB}}$ = lower bound of the predicted retention time
		
		\ENSURE  Retention time for an STT-MRAM cell
		\STATE Initialization
		\FOR {$i = 0$ to $N_{\mathrm{seh}}-1$}
		\STATE $t_{\mathrm{ret}}[i]=1/2\times(t_{\mathrm{UB}}+t_{\mathrm{LB}})$
		\FOR {$j=0$ to $N-1$}
		\STATE Regular write of a $test \: pattern$
		\STATE Wait for time $t_{\mathrm{ret}}[i]$
		\STATE Regular read for a $readout \: data$
		\IF{$readout \: data \ne test \: pattern$}
		\STATE   $t_{\mathrm{UB}}=t_{\mathrm{ret}}[i]$
		\ELSE
		\STATE  $t_{\mathrm{LB}}=t_{\mathrm{ret}}[i]$
		\ENDIF
		\STATE $t_{\mathrm{ret}}[i]=1/2\times(t_{\mathrm{UB}}+t_{\mathrm{LB}})$
		\ENDFOR
		
		\STATE $T_{\mathrm{ret}}=\frac {\sum t_{\mathrm{ret}}[i]}{N_{\mathrm{seh}}}$
		\ENDFOR
		\RETURN $T_{\mathrm{ret}}$
	\end{algorithmic}
\end{algorithm}

\subsubsection{Burn-in Methods}
To reduce the prohibitive test time of retention time in STT-MRAMs, Burn-in test techniques to compress the retention time is an effective way. Since the thermal stability $\Delta$ of the MTJ device significantly depends on  ambient conditions such temperature, magnetic field, and disturb current, it is viable to change those conditions to compress the $\Delta$ value, thus accelerating the process of  retention fault.  Combining Equations (\ref{eq:Delta}, \ref{eq:Delta_offsetfield},  \ref{eq:Delta_I}), we derive the following equation:
\begin{equation}
   \Delta(T,I,H_{\mathrm{offset}})=\frac {E_\mathrm{B}} {k_\mathrm{B}T}(1-\frac {I} {I_{\mathrm{c}0}})(1-\frac {H_{\mathrm{offset}}} {H_\mathrm{k}})^2. \label{eq: Delta_compression}
\end{equation}
Equation (\ref{eq: Delta_compression}) indicates that an increase in temperature $T$ leads to a smaller $\Delta$ value. Moreover, The $\Delta$ value decreases with a current $I$ flowing through the MTJ device and an external offset field $H_{\mathrm{offset}}$ opposite to the intrinsic anisotropy field $H_\mathrm{k}$ of the FL. 
 Ghosh et al. \cite{Iyengar2016} proposed two algorithms to test the retention time of STT-MRAM cells with a thermal burn-in scheme, whereby the retention time can be $1000\times$ smaller when the temperature rises to \SI{125}{\celsius}.

The first algorithm is named as binary-search-based retention test, as depicted in Algorithm 2.  This algorithm describes the process of testing the retention time ($T_{\mathrm{ret}}$) of a single bit-cell. Initially, a lower bound search time ($t_{\mathrm{LB}}$) and an upper bound  search time ($t_\mathrm{UB}$) for the retention time are  selected according to the predicted thermal stability of STT-MRAM arrays under test. Thereafter a test data pattern is written into the first memory cell under test, followed by a fixed period of wait time ($t_{\mathrm{ret}}$), equal to the mean  of $t_{\mathrm{LB}}$ and $t_\mathrm{UB}$. After the wait time, the data in this cell is read out and compared with the original test pattern.  If they are the same, meaning no retention fault occurs, the $t_{\mathrm{LB}}$ value is updated with the wait time $t_{\mathrm{ret}}$. If not,  the $t_\mathrm{UB}$ value is updated with the $t_{\mathrm{ret}}$. As this process  iterates, the $t_{\mathrm{ret}}$ value gradually approaches the actual retention time of the cell under test.  Depending on the accuracy we desire, the overall test time increases with the number of binary searches for averaging and the iteration cycles in each search.

The second algorithm searches the retention time in a linear way, described below in Algorithm 3. With this algorithm, the  time step ($T_{\mathrm{step}}$) in each search and the number of searches ($N_{\mathrm{seh}}$) are given to determine the search resolution and test accuracy, respectively. The test starts with a regular write  of a given test data pattern into the cell under test. Then it periodically reads the data back with the time step $T_{\mathrm{step}}$ to compare it with the original data pattern. This search process proceeds until a mismatch is observed, similar to the polling scheme in CPU. Clearly, a small time step results in a high test accuracy at the expense of more read operations.

 \begin{algorithm}
 \caption{ Burn-in retention test based on linear search.}
 \begin{algorithmic}
 \renewcommand{\algorithmicrequire}{\textbf{Input:}}
 \renewcommand{\algorithmicensure}{\textbf{Output:}}
 \REQUIRE $N_{\mathrm{seh}}$ = the number of linear searches for averaging
 \REQUIRE  $T_{\mathrm{step}}$ = resolution of the retention time test
 \ENSURE  Retention time for an STT-MRAM cell
 \STATE Initialization
         \FOR {$i = 0$ to $N_{\mathrm{seh}}-1$}
                \STATE Regular write of a $test \: pattern$
                \STATE Reset $step \:  counter$
                \WHILE{$readout \: data = test \: pattern$}
                        \STATE Wait for time $T_{\mathrm{step}}$
                        \STATE $step \:  counter$++
                        \STATE Regular read for a $readout \: data$
                        \STATE $t_{\mathrm{ret}}[i]$=$T_{\mathrm{step}}\times step \: counter$
                \ENDWHILE
         \ENDFOR
         \STATE $T_{\mathrm{ret}}=\frac {\sum t_{\mathrm{ret}}[i]}{N_{\mathrm{seh}}}$
 \RETURN $T_{\mathrm{ret}}$
 \end{algorithmic}
 \end{algorithm}

\subsubsection{Algorithm Comparison}
 As aforementioned, the statistical method based on the injection of a weak disturb current in Algorithm 1 is extremely time-consuming, due to the low occurrence rate of the retention fault at room temperature. Algorithm 2 and 3 can be superior to Algorithm 1 if the compression rate is very high such that the compressed retention time is on the order of ms or below. This is applicable to STT-MRAMs with a low $\Delta$ for cache applications, where the averaged block lifetime is below \SI{1}{\second}. However, for SCM applications requiring 10+ years of retention time, it is of no avail to compress and test the retention time with a linear search or a binary search. For instance, compared to \SI{0.5}{\second} for testing a bit-cell with $\Delta=60$ using Algorithm 1, Algorithm 2 or 3 takes more than $10\times$ test time  even with a compression rate as high as $10^8$ combining both the thermal burn-in and the disturb current techniques, as claimed in \cite{Iyengar2016}. Under such a high compression rate, the resulted retention time is approximately \SI{3}{\second}, which is the minimum test time for a bit-cell using Algorithm 2 or 3. Therefore, there is still  space to explore for  testing the retention time of STT-MRAMs in a cost-efficent way for various applications. One possible solution is to combine the following three burn-in techniques to aggressively compress the thermal stability $\Delta$: 1) thermal burn-in with an elevated temperature; 2) electrical burn-in with the injection of a weak disturb current; 3) magnetic burn-in with an external magnetic field $H_{\mathrm{offset}}$ opposite to the intrinsic anisotropy field $H_\mathrm{k}$ of the FL.

\section{DESIGN FOR TESTABILITY}
As aforementioned, permanent fault models such as SAF and TF in STT-MRAMs can be detected by March tests. However, March tests cannot guarantee the detection of STT-MRAM transient faults (i.e., TWF, TRDF, and RTF) which only take place in some specific cycles with certain occurrence rates. Thus, transient faults require \textit{design-for-testability} (DfT) techniques for detecting them. In this section, we will discuss two circuit designs proposed in the literature for the detection of TRDF and RTF. 
\label{sec:dft}

\begin{figure}[!t]
    \centering
    \subfloat[Illustration of the abrupt  current increase  due to the occurance of TRDF   while reading 1(AP) state. Reprinted from \cite{Bishnoi2015}.]{\includegraphics[width=0.4 \textwidth]{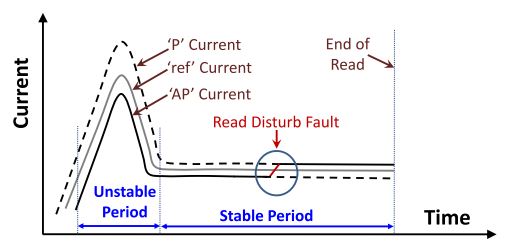}%
    \label{fig_TRDF_Principle}}
    \hfil
    \subfloat[Bit-flip detection circuit by means of comparing the read current ($I_{\mathrm{rd}}$) going through the cell under read and the current ($I_{\mathrm{ref}}$) flowing through the reference cell \cite{Bishnoi2015}.]{\includegraphics[width=0.4 \textwidth]{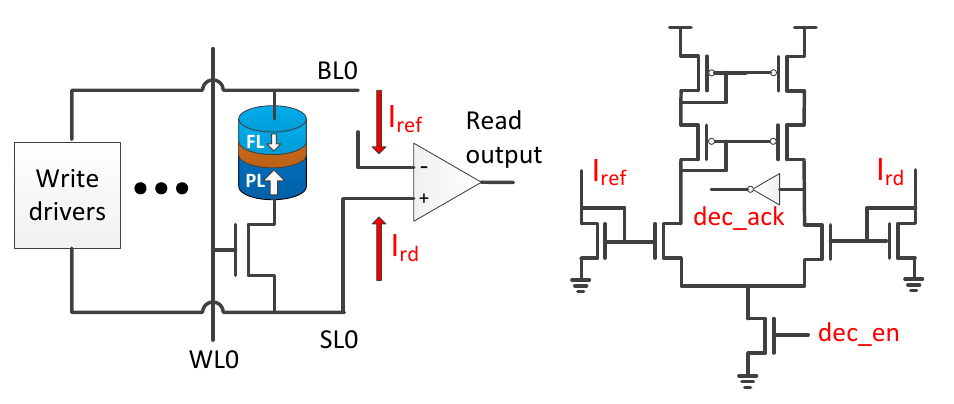}%
    \label{fig_TRDF_Detection_Circuit}}
    \caption{Design for TRDF detection: (a) principle, and (b) circuit design.}
    \label{fig_TRDF_Detection}
\end{figure}

\subsection{ Design for TRDF Detection }
\label{subsec:design4TRDF}
Transient read disturb fault (TRDF) in STT-MRAMs is a transient fault and is increasingly becoming a threat to the reliability of STT-MRAMs, as we discussed previously. Thus, developing DfT designs to  facilitate TRDF detection has received great attention. Two similar circuit-level DfT techniques have been proposed separately in \cite{Bishnoi2015,Ran2016} to detect TRDF. Both of them are based on a key observation: TRDF changes the MTJ resistance of the victim cell, which in turn affects the current amplitude during read operations.

Fig. \ref{fig_TRDF_Detection}\subref{fig_TRDF_Principle} illustrates that  the read current  goes up abruptly from below $I_{\mathrm{ref}}$ to above $I_{\mathrm{ref}}$ when a TRDF occurs during a read 1(AP) operation. The solid gray line indicates the current  flowing through the reference cell during read operations, and the dashed lines below and above the solid gray line indicate the currents flowing through fault-free cells in AP state and P state, respectively. The solid black line indicates the actual current change over time in a read operation during which a TRDF (marked with the blue circle) takes place. This observation is leveraged to  detect TRDF by integrating a dedicated circuit into the sense amplifier to track the current change  in read operations. Fig. \ref{fig_TRDF_Detection}\subref{fig_TRDF_Detection_Circuit} shows the bit-flip detection circuit design. The $dec\_en$ signal is only asserted when reading 1(AP) state.  In this case, two current mirrors are used to copy the current $I_{\mathrm{rd}}$ going through the cell under read and the current $I_{\mathrm{ref}}$ flowing through the reference cell from the inputs of  the sense amplifier. As long as  $I_{\mathrm{rd}}$ is smaller than $I_{\mathrm{ref}}$, the disturb acknowledgment signal $dec\_ack$ remains at ``0'' state. However, if $I_{\mathrm{rd}}$ increases abruptly above $I_{\mathrm{ref}}$ (i.e., TRDF occurs) in the stable period of the sensing process, the $dec\_ack$ signal immediately makes a transition to ``1'' state, indicating a detection of TRDF. Despite the effectiveness of detecting TRDFs by means of this technique at the expense of negligible power and area overhead, neither of the two papers provides further insights into how to repair or tolerate TRDF in STT-MRAMs.

\begin{figure}[!t]
    \centering
    \subfloat[Repeatedly applying weak write currents to 16 rows of cells  simultaneously in test mode to get statistical estimation of $Pr (I_{\mathrm{wwr}})$.]{\includegraphics[width=0.4 \textwidth]{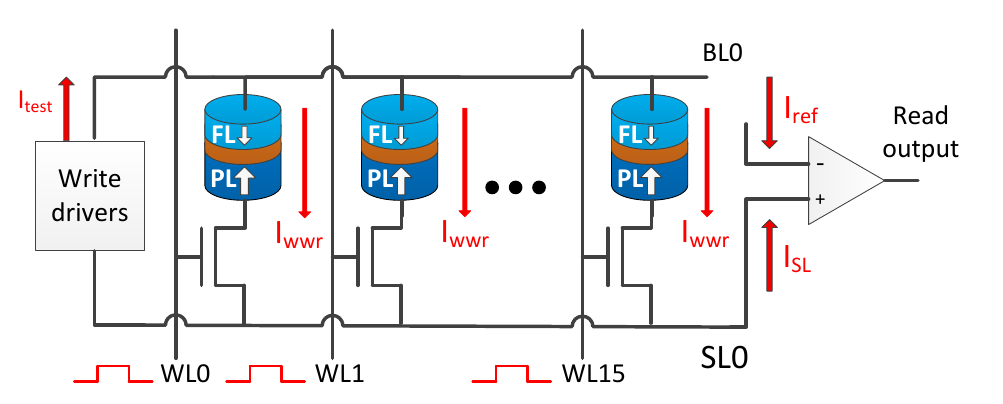}%
    \label{fig_Error_Detection}}
    \hfil
    \subfloat[Tracking the current on the SL ($I_{\mathrm{SL}}$) during tests so as to avoid unnecessary read operations when no bit flip  occurs \cite{Insik_ITC_2016}.]{\includegraphics[width=0.4 \textwidth]{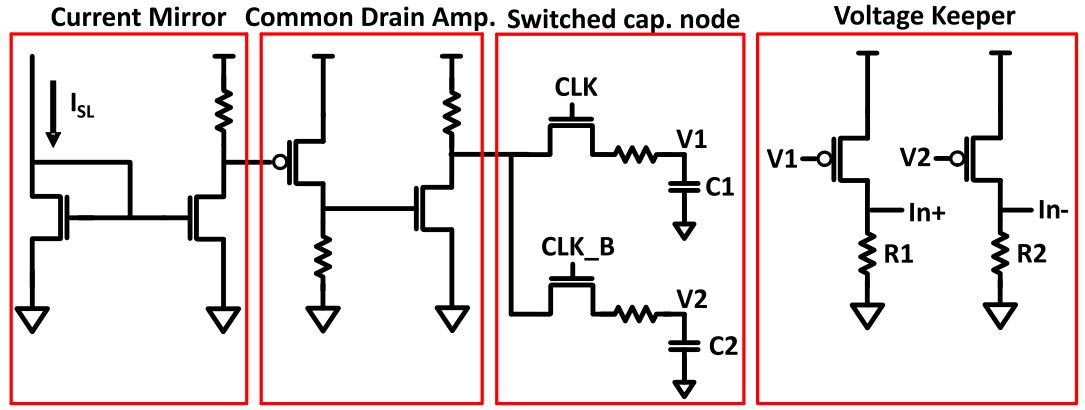}%
    \label{fig_EDcircuit}}
    \caption{Retention test implementation: (a) measurement of read disturb probability $Pr (I_{\mathrm{wwr}})$, and (b) bit-flip detection circuit. }
    \label{fig_ErrorDetection_Circuit}
\end{figure}

\subsection{ Design for Retention Test }
\label{subsec:design4retentiontest}
In order to test the retention time  of STT-MRAM cells more efficiently, Yoon et al. improved Algorithm 1 and  implemented a memory build-in-self-test (MBIST) \cite{Insik_ITC_2016,Yoon_QED_2017}. As aforementioned, Algorithm 1 has many limitations including: 1) the retention test has to be carried out in an operating region where  the switching probability $Pr(I_{\mathrm{wwr}})$ is very small under a small current $I_{\mathrm{wwr}}$; 2) the majority of  read operations after applying the weak write current are not necessary due to the small switching probability; 3) the test time is prohibitive and increases with  $\Delta$ value and array size. To overcome these limitations, the MBIST implementation leverages the philosophy behind the preceding TRDF detection to avoid unnecessary read operations when the retention fault does not occur. In addition, applying the weak write current to multiple rows simultaneously instead of a single row in Algorithm 1 within a test iteration significantly speeds up the test process.

The retention test process starts with writing a predefined data pattern into the cells under test. The data pattern is set to 1 due to two considerations. First, the small current disturbance is uni-directional (i.e., from AP state to P state).  Second, the thermal stability is  at the lowest value when the cell is in AP state due to the inter-cell magnetic coupling among neighboring cells  as we discussed in Section \ref{subsec:magneticcoupling}.  Thereafter, a weak write current ($I_{\mathrm{wwr}}$) is applied to 16 rows of cells at the same time, as illustrated in Fig. \ref{fig_ErrorDetection_Circuit}\subref{fig_Error_Detection}. By tracking the current change on the SL  using a dedicated bit-flip detection circuit (see Fig.  \ref{fig_ErrorDetection_Circuit} \subref{fig_EDcircuit}), a bit-flip can be immediately detected as $I_{\mathrm{SL}}$ increases slightly when any cell among the 16 cells flips from AP state to P state. The slight increase  in $I_{\mathrm{SL}}$ is first amplified by a current mirror and transferred to voltage difference, which is further amplified by a multi-stage common drain amplifier. Thereafter, the switched capacitors C1 and C2 sample the voltage alternatively based on the CLK and CLK\_B signals. At the end of this chain, a voltage keeper is used to make sure the voltage difference between the In+ and In- nodes  is always higher than \SI{10}{\milli\volt} to avoid metastability. When the sense amplifier is enabled, the voltage difference between the In+ and In- nodes is fully amplified to VDD and GND.

 With this  scheme for the  retention test, experimental results in \cite{Insik_ITC_2016} show a 93.75\% improvement in test time compared to the  conventional test scheme based on Algorithm 1.
\section {Discussion}
\label{sec:discussion}
As one of the most promising NVM technologies, STT-MRAM offers competitive write performance, endurance and data retention.  The  tunability of these three aspects also makes it customizable for a variety of applications such as last-level caches, Internet-of-Things, and automotive. Therefore, STT-MRAM has received a large amount of investment from major semiconductor companies including Intel and Samsung, which have demonstrated its manufacturability in recent years.
However, several challenges remain to be addressed before its mass production at different levels, including failure mechanisms, fault modeling, and test development. In this section, we will discuss the key challenges at these three levels. 
\begin{figure}[!t]
	\centering
	\includegraphics[width=0.5 \textwidth ]{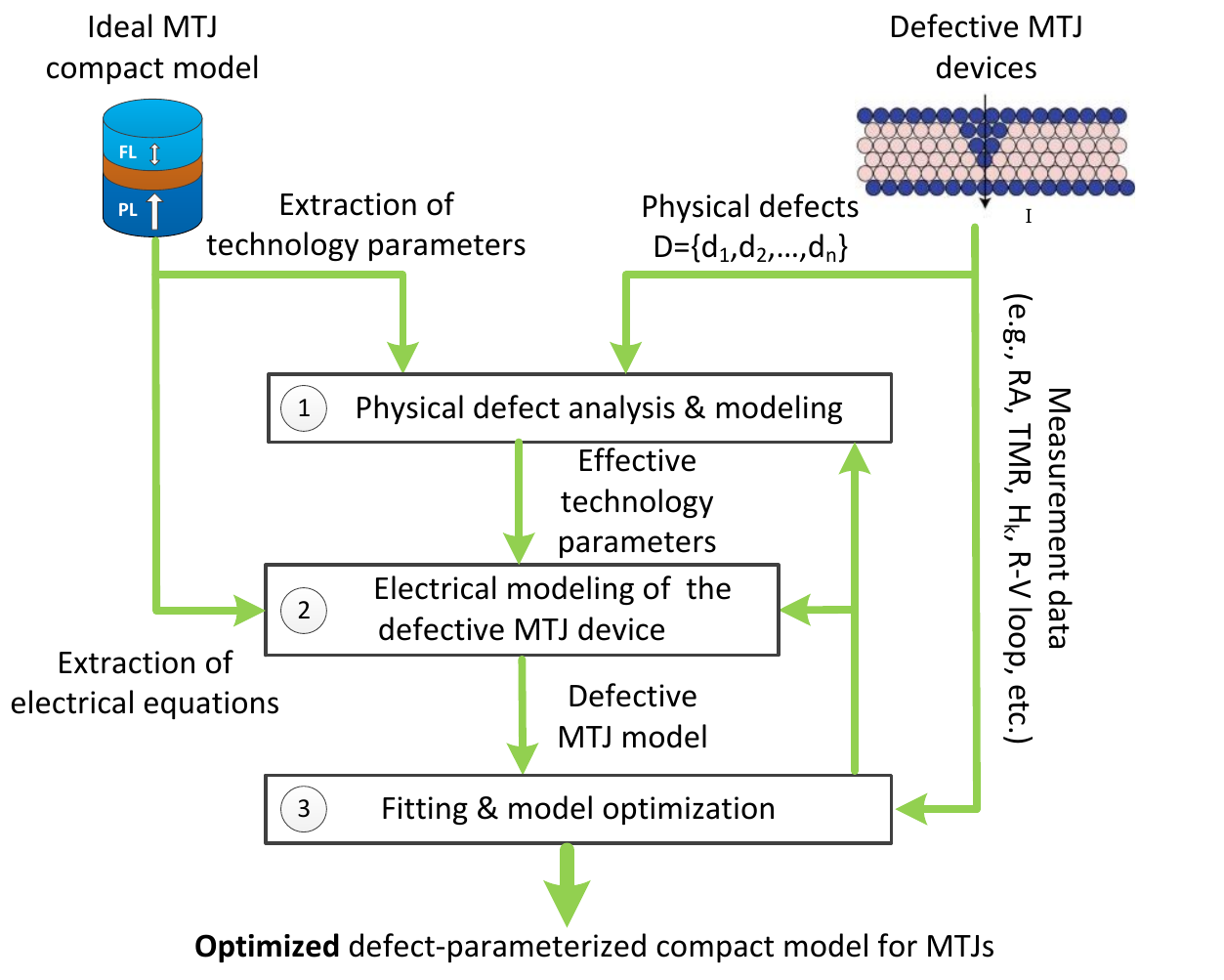}
	\caption{Electrical modeling flow for STT-MRAM-specific defects. }
	\label{fig_methodology}
\end{figure}

\subsection{Challenges for Failure Mechanisms}
\label{subsec:challenges4Failuremechanisms}
In Section \ref{sec:defect_mechanisms}, we have discussed the main  failure mechanisms for STT-MRAM, namely manufacturing defects, extreme process variations, magnetic coupling, STT stochastic switching, and thermal fluctuation.  Next, we discuss the remaining challenges for these failure mechanisms.

First,  the cause, location, occurrence rate, and electrical impact of STT-MRAM manufacturing defects have not been fully explored yet. We have collected and presented in this survey all potential manufacturing defects reported in the literature especially for those related to the fabrication of MTJ devices which are the data-storing elements for STT-MRAMs. However, more research work needs to be done in terms of investigating the characteristics of each defect in STT-MRAMs. For example, physical measurement (e.g., TEM) of the defect sheds light on  the cause and location; this is important to understand how and where the defect is introduced. In addition, investigating the occurrence rate of each defect indicates how important the defect is. Last but not least, electrical characterization of defects especially those occurred in the MTJ device is also  crucial to understand the electrical consequence at memory-cell level.

Second, developing accurate models for STT-MRAM-specific defects is still a challenge.  Conventionally, physical defects are modeled as resistive shorts and opens which are used to develop fault models based on circuit simulations. Although this defect modeling approach is still qualified to model defects in interconnects of STT-MRAMs, it may not  be applicable for defects in MTJ devices. In our recent work \cite{Wu2019a}, we demonstrated with silicon measurement data that using linear resistors to model pinhole defects in the MgO tunnel barrier of MTJ devices is inaccurate. This is due to the fact that  linear resistors cannot capture  defect-induced changes in magnetic properties, which are as important as electrical ones for MTJ devices. Furthermore, inaccurate defect modeling may result in incorrect fault models, leading to low-quality test solutions for non-existing problem. Therefore, it is paramount to understand the physics of STT-MRAM-specific defects so as to provide accurate defect models. This is important to guarantee a high-quality test solution as well as to improve the manufacturing process itself so as to improve yield. In \cite{Wu2018}, we proposed a three-step defect modeling methodology, where the effects of physical defects on the MTJ device are incorporated into the  technology parameters of the MTJ device and thereafter on its  electrical parameters, as shown in Fig.~\ref{fig_methodology}. It is worth-noting that measurement data of real  defective MTJ devices is a vital to ensure the accuracy  of the defect models for a specific STT-MRAM design and manufacturing process.

Third, with technology downscaling, the unique failure mechanisms in STT-MRAMs such as magnetic coupling, STT stochastic switching, and thermal fluctuation are increasingly become reliability issues. As  discussed previously, these new failure mechanisms result in transient faults (e.g., TWF, TRDF) which  intermittently appear in some specific cycles during usage.  This has been one of the  major obstacles limiting the mass commercialization of STT-MRAM in the industry. To reach desired write error rate and read error rate meeting the industry's standards, mitigation techniques are required at device, circuit, and system levels. For example, at device level, the double SAF structure of  MTJ stack  was reported to be effective in compensating the offset field $H_{\mathrm{offset}}$ at the FL \cite{Augustine2011,Han2015}. In addition, innovations on MTJ stack designs to further cut down the switching current below \SI{100}{\micro\ampere} while maintaining a fast switch speed for sub-\SI{10}{\nano\second} are also urgently needed \cite{Saida2016}. At circuit level, a large write margin (i.e., higher write current amplitude and duration) can be employed to ensure a high switching probability of the state in MTJ devices from a circuit design perspective.  Another example of circuit-level techniques is write-verify-write scheme which has been adopted by Intel in \cite{Golonzka2018}. In this scheme, a read operation is applied immediately following a write operation to verify whether or not the write operation is successful. If failed, a second write operation is required to write the same data again in order to guarantee the data has been successfully written to the addressed cell. However, these two circuit-level techniques  come at the expense of higher energy consumption and performance loss.  At system level, error correction codes (ECCs) are commonly used to tolerate  some bits of transient faults at run time by introducing redundant parity bits. Nevertheless, strong ECCs are also costly, as the area and latency overhead greatly increases with the number of correction bits.  


\subsection{Challenges for Fault Modeling}
\label{subsec:challenges4faultmodeling}
As introduced in Section \ref{sec:fault_models}, permanent faults in STT-MRAMs are induced by manufacturing defects and extreme process variations while transient faults are caused by magnetic coupling, STT stochastic switching, and thermal fluctuation. Both permanent and transient faults are the high-level representation of failure mechanisms. Thus, it is of great importance to develop accurate fault models reflecting the physical natures of above-mentioned  failure mechanisms.  Next, we discuss challenges in terms of  fault modeling for STT-MRAMs. 

First, a consistent fault nomenclature is necessary to unify all existing fault terminologies. At present, the names of fault models for STT-MRAMs in the literature are chaotic and ambiguous. This can be  reflected at three aspects as follows. 
\begin{itemize}
	\item Current fault model names are not intuitively interpretable. For instance, Vatajelu et al. \cite{Vatajelu2017} defined the transient write fault caused by the  STT stochastic switching property as undefined write fault (UWF).  However, this name does not imply the transient attribute meaning that  this fault only occurs in some cycles with certain probability during run time. Furthermore, the term ``undefined'' here is rather ambiguous, as UWF in RRAM means that the cell resistance is settled to an intermediate (undefined) state between logic 0 and 1. 
	\item The same fault model name describes different faulty behaviors. For example, TRDF differs from RDF in terms of physical causes, repeatability, and readout data, etc., as shown in Table \ref{table:RDF_TRDF}.  Moreover, RDF is a type of permanent faults which should be targeted during during manufacturing tests, while TRDF is transient thus needs to be excluded during manufacturing tests. However, these two different faults are both referred to as read disturb/destructive fault in the literature. 
	\item The same faulty behavior is named differently among different researchers. As discussed in Section \ref{sec:fault_models}.B, transition fault is also named as slow write fault, and write disturb coupling fault is named as coupling fault or write disturb fault by different researchers. The  chaos in terminology usage may impede the communication and collaboration in the test community.
\end{itemize}

Second, a systematic fault analysis methodology is required. Conventionally, fault analysis is based on circuit simulation with the injection of resistive defects (i.e., opens and shorts);  the defect strength  is determined by the resistance of the injected linear resistors. Thereafter, fault models are proposed to describe the resultant faulty behaviors. This method however has limitations as follows.
\begin{itemize}
	\item Some of the fault models developed with the conventional linear-resistor-based defect modeling approach may not exist in practice. As afore-mentioned, defects in MTJ devices require a more accurate defect modeling approach to capture the changes in both the device's magnetic and electrical properties. Using linear resistors to model those defects may results in wrong fault models which do not exist in practice, leading to an inefficient test solution.
	\item  The current expression of faulty behaviors is not rigorous and systematic. As it can be observed in the papers related to STT-MRAM testing, memory fault behaviors are described by fault models; their names indicate the corresponding fault behaviors. This fault description method is not easily interpretable from the name itselt, and it can be even confusing sometimes. One example is the previously-discussed undefined write fault UWF where the exact fault behavior is not very clear to a reader.  In our latest paper \cite{Wu2019_TETC}, we defined  the complete fault space and nomenclature,  qualified to describe all faults in STT-MRAMs.
	\item The fault analysis  is not complete and systematic. As discussed in Section \ref{subsec:static_faults}, the state-of-the-art of fault modeling for STT-MRAMs is limited to static faults in STT-MRAM cell array. However, memory cells with defects which do not sensitize any static faults are not necessarily fault-free. Dynamic faults \cite{hamdioui2003dynamic} which are sensitized by more than one operations should also be considered in the fault analysis. Furthermore, even if a small defect does not sensitize any static faults or  dynamic faults,  it may cause weak faults, leading to a shorter lifetime or higher in-field failure rate. Thus, attention should also be paid to those weak faults especially when it comes to defective-part-per-billion (DPPB) level test requirement. Moreover, faults located in address decoders and read\&write circuits are also need to be covered, since these peripheral circuits are indispensable for STT-MRAM chips but may also subject to manufacturing defects.  
\end{itemize}

\subsection{Challenges for Test/DfT Development}
\label{subsec:challenges4testdevelopment}
Test algorithms and DfT designs targeting STT-MRAMs are still not  established yet. Research in this field is ongoing to guarantee a high percentage of defect/fault coverage meanwhile minimizing the test cost and time. For strong faults (including static and dynamic faults) which can be sensitized by a sequence of read/write operations, March tests are typically used to detect them. For weak faults, DfT designs are needed to guarantee the detection. To ensure a high-quality test solution for STT-MRAMs, the following issues need to be addressed.

First, an effective and efficient March test necessitates accurate fault models. Since fault models are the targets of March tests, unrealistic fault models may result in a poor-quality March test for non-existing problem, meaning a waste of time time. Worse still, test escape can happen when manufacturing defects are not well modeled and detected by the test. 

Second, transient faults should be excluded somehow during manufacturing tests. As discussed in Section \ref{subsec:static_faults}, transient faults are mainly caused by  magnetic coupling, STT stochastic switching, and thermal fluctuation. They do not appear in every cycle and can be self-corrected or tolerated by circuit/system level techniques such as ECCs. Therefore, manufacturing tests should target permanent faults rather than transient faults. However, given the high requirement on the time and cost for mass production tests, transient faults may show up in good devices during tests, causing overkilling  good chips. This is known as yield loss. Therefore, it is of great importance to distinguish transient faults from permanent faults during manufacturing tests to avoid failing good chips. 

Third, STT-MRAM retention  test is extremely time consuming especially for those chips for storage applications requiring a retention time of $>$$10$ years. This makes integrating retention test to production tests (where every fabricated chip is tested) prohibitively expensive, similar to other burn-in/stress tests. In Section \ref{subsec:retentiontest} and \ref{subsec:design4retentiontest},  three test algorithms along with a DfT design for STT-MRAM retention test have been introduced, but none of them is as cost-efficient as desired in practice. To further reduce the retention test time, stress tests such as magnetic burn-in, thermal burn-in, and disturb current application, along with DfT designs provide a feasible solution. 

\section {CONCLUSION}
\label{sec:conclusion}
This paper surveys the state-of-the-art on STT-MRAM testing at three abstraction levels, namely failure mechanisms, fault models, and tests. We presents that manufacturing defects, extreme process variations, magnetic coupling, STT-switching stochasticity, and thermal fluctuation are the main failure mechanisms for STT-MRAMs. To ensure a good test solution, it is paramount to understand these failure mechanisms so as to propose accurate model for them. Special attentions should be paid to the defects in  MTJ devices which are the data-storing elements in STT-MRAMs. Rather than modeling those MTJ-related defects as  linear resistors conventionally, it is necessary to open the black box of the MTJ model and incorporate the physical defects into the device model. Accurate fault modeling requires both accurate defect models and a systematic fault analysis methodology. The proposed test algorithms and DfT designs in the literature are very limited and the effectiveness is still in doubt. A cost-efficient March test is required to detect permanent faults while excluding transient faults. Meanwhile, DfT designs or stress tests such as magnetic and thermal burn-in are also need to be taken into account to detect weak faults especially when considering high test requirements toward DPPB level.  As STT-MRAM testing is still an emerging research topic, manufacturing tests are  far from well established yet.  Therefore, more research work needs to be done on this topic to address remaining challenges at these three abstraction levels.


\ifCLASSOPTIONcaptionsoff
\newpage
\fi

\def\url#1{}
\bibliographystyle{IEEEtran}
\bibliography{library}{}

\end{document}